\documentclass[fleqn,usenatbib]{mnras}

\usepackage{mathptmx}
\usepackage{booktabs}

\usepackage[T1]{fontenc}
\usepackage[dvipsnames]{xcolor}

\DeclareRobustCommand{\VAN}[3]{#2}
\let\VANthebibliography\thebibliography
\def\thebibliography{\DeclareRobustCommand{\VAN}[3]{##3}\VANthebibliography}

\usepackage{graphicx}	
\usepackage{amsmath}	
\usepackage{txfonts}
\usepackage{subeqnarray}
\usepackage{caption}
\usepackage{subcaption}
\usepackage{url}
\usepackage{threeparttable}
\usepackage{multirow}

\newcommand{\kms}{km.s$^{-1}$}

\newcommand{\kp}{K$_{\rm{p}}$}
\newcommand{\vsys}{V$_{\rm{sys}}$}

\newcommand{\hdb}{HD\,189733\,b}

\newcommand{\GLA}{Gl\,15 A\ }

\defcitealias{Maciejewski2014}{Ma14}
\defcitealias{rosenthal2021}{Ro21}
\defcitealias{cutri2003}{Cu03}
\defcitealias{stassun2017}{St17}
\defcitealias{addison2019}{Ad19}
\defcitealias{bourrier2018}{Bo18}
\defcitealias{baluev2015}{Ba15}
\defcitealias{trifonov2021}{Tr21}
\defcitealias{claret2011}{Cl11}
\defcitealias{fouque2018}{Fo18}
\defcitealias{turner2016}{Tu16}

\newcommand{\yes}[1]{{\color{Green}{{#1}}}}

\title[ATMOSPHERIX: atmosphere characterization]{ATMOSPHERIX: II- Characterising exoplanet atmospheres through transmission spectroscopy with SPIRou}

\author[Debras, F. et al.]{
Florian Debras,$^{1}$\thanks{E-mail: florian.debras@irap.omp.eu}
Baptiste Klein,$^{1,2}$
Jean-François Donati, $^{1}$
Thea Hood,$^{1}$
Claire Moutou,$^{1}$
Andres Carmona, $^{3}$\newauthor
Benjamin Charnay,$^{4}$
Bruno Bézard, $^{4}$
Pascal Fouqué, $^{1}$
Adrien Masson,$^{4}$
Sandrine Vinatier,$^{4}$
Clément Baruteau, $^{1}$\newauthor
Isabelle Boisse,$^{5}$
Xavier Bonfils,$^{3}$
Andrea Chiavassa,$^{6}$
Xavier Delfosse,$^{3}$
Guillaume Hebrard,$^{7,8}$\newauthor
J\'er\'emy Leconte,$^{9}$
Eder Martioli,$^{7,10}$
Merwan Ould-elkhim,$^{1}$
Vivien Parmentier, $^{2,6}$
Pascal Petit,$^{1}$\newauthor
William Pluriel,$^{11}$
Franck Selsis,$^{9}$
Lucas Teinturier,$^{4,13}$
Pascal Tremblin,$^{12}$
Martin Turbet, $^{13,9}$
Olivia Venot,$^{14}$ \newauthor
and Aurélien Wyttenbach$^{3}$
\\
$^{1}$IRAP, Université de Toulouse, CNRS, UPS, Toulouse, France\\
$^{2}$Department of Physics, University of Oxford, OX13RH, Oxford, UK\\
$^{3}$Université Grenoble Alpes, CNRS, IPAG, 38000 Grenoble, France\\
$^{4}$LESIA, Observatoire de Paris, Université PSL, Sorbonne Université, Université Paris Cité, CNRS, 5 place Jules Janssen, 92195 Meudon, France\\
$^{5}$Aix Marseille Universite, CNRS, Laboratoire d’Astrophysique de Marseille UMR 7326, 13388, Marseille, France\\
$^{6}$Université Côte d'Azur, Observatoire de la Côte d'Azur, CNRS, Lagrange, CS 34229, Nice, France \\
$^{7}$Institut d'astrophysique de Paris, UMR7095 CNRS, Universit\'e Pierre \& Marie Curie, 98bis boulevard Arago, 75014 Paris, France\\
$^{8}$Observatoire de Haute-Provence, CNRS, Universit\'e d'Aix-Marseille, 04870 Saint-Michel-l'Observatoire, France\\
$^{9}$Laboratoire d'astrophysique de Bordeaux, Univ. Bordeaux, CNRS, B18N, all\'ee Geoffroy Saint-Hilaire, 33615 Pessac, France \\
$^{10}$Laborat\'{o}rio Nacional de Astrof\'{i}sica, Rua EstadosUnidos 154, 37504-364, Itajub\'{a} - MG, Brazil \\
$^{11}$Observatoire astronomique de l'Universit\'e de Gen\`eve, chemin Pegasi 51, 1290 Versoix, Switzerland \\
$^{12}$Universite Paris-Saclay, UVSQ, CNRS, CEA, Maison de la Simulation, 91191, Gif-sur-Yvette, France\\
$^{13}$Laboratoire de Météorologie Dynamique/IPSL, CNRS, Sorbonne Université, Ecole Normale Supérieure, Université PSL, Ecole Polytechnique,\\
Institut Polytechnique de Paris, 75005 Paris, France \\
$^{14}$Université de Paris Cité and Univ Paris Est Creteil, CNRS, LISA, F-75013 Paris, France\\
}

\date{Accepted XXX. Received YYY; in original form ZZZ}

\pubyear{2022}

\begin{document}
\label{firstpage}
\pagerange{\pageref{firstpage}--\pageref{lastpage}}
\maketitle

\begin{abstract} In a companion paper, we introduced a publicly-available pipeline to characterise exoplanet atmospheres through high-resolution spectroscopy. In this paper, we use this pipeline to study the biases and degeneracies that arise in atmospheric characterisation of exoplanets in near-infrared ground-based transmission spectroscopy. We inject synthetic planetary transits into sequences of SPIRou spectra of the well known M dwarf star \GLA, and study the effects of different assumptions on the retrieval. We focus on (i) mass and radius uncertainties, (ii) non isothermal vertical profiles and (iii) identification  and retrieval of multiple species. We show that the uncertainties on mass and radius should be accounted for in retrievals and that depth-dependent temperature information can be derived from  high-resolution transmission spectroscopy data. Finally, we discuss the impact of selecting wavelength orders in the retrieval and the issues that arise when trying to identify a single species in a multi-species atmospheric model. This analysis allows us to understand better the results obtained through transmission spectroscopy and their limitations in preparation to the analysis of actual SPIRou data.  
\end{abstract}

\begin{keywords}
exoplanets -- planets and satellites: atmospheres -- planets and satellites: gaseous planets -- techniques: spectroscopic -- methods: data analysis
\end{keywords}



\section{Introduction}\label{sec:intro}

In a companion paper (Klein et al., submitted to MNRAS, hereafter named paper I), we have introduced our publicly-available pipeline for the analysis of high resolution spectroscopy (HRS) data of exoplanet atmospheres. This pipeline was developed in the framework of the ATMOSPHERIX consortium, a gathering of observers and theoreticians created to optimize the study of ground-based HRS for exoplanet atmospheres at the French level. We have shown the validity and robustness of this pipeline for single-component isothermal planetary atmospheres. However, we know this is a crude simplification as more and more molecular species are discovered in exoplanet atmospheres \citep[see review in][]{Guillot2022} and departures from vertically-isothermal atmospheres are also commounly found thanks to stronger temperature constraints \citep[e.g.][]{Haynes2015,Gibson2020}.

More complex models would therefore be needed to be representative of actual observations but the more complex the model, the more degenerate the problem. It is therefore an important task to understand the sources of degeneracies in atmospheric retrievals in order to provide the most reliable parameter estimates.
Such degeneracies have already been studied in low resolution spectroscopy (LRS) for more than 20 years (see e.g., \citet{Brown2001} and the references in the introduction of \citet{Welbanks2019}) but are less extensively studied in HRS, particularly in the infrared.  \citet{Fisher2019} have studied the information that can be obtained through the sodium doublet in the visible, concluding that HRS alone is not enough to determine appropriately the pressure that are probed by the sodium lines. The combination with LRS in \citet{Pino2018} might allow to resolve some of these questions.  We obtained similar conclusions when including clouds in our companion paper, where the loss of the continuum by HRS exacerbated a degeneracy between cloud top pressure and water content. Clouds, in general, are a major source of work to understand plausible degeneracies in the spectra, both at high and low resolution (see e.g., \citet{kitzmann2018,Barstow2020}).  Inclusion of multi-dimensional effects further complicates this picture \citep{Line2016,Pluriel2020,Welbanks2022}.

In this paper, we therefore focus on a few degeneracies and potential biases that are inherent to HRS with application on synthetic SPIRou transit data. We test three cases: uncertainties in the planet's mass and radius, non isothermal vertical structures and models with multiple molecular species with comparable mixing ratios. We first recall the process of data generation and reduction in section \ref{sec:Data}. In section \ref{sec:results}, we then present our test cases and the results of template matching and Bayesian retrieval on the atmospheric parameters. This leads us to discuss how to optimize the detection and the ways forward in section \ref{sec:discussion}, before concluding in Section~\ref{sec:conclusion}.

\begin{table*}
\centering
\begin{threeparttable}
\begin{tabular}{cccc}
  \hline
  \textbf{Stellar parameters} & \multicolumn{3}{c|}{\textbf{Gl 15A}}  \\
   & \multicolumn{2}{c|}{Value} & Reference \\ 
  \hline
  Mass ($M_\odot$) & \multicolumn{2}{c|}{0.400 $\pm$ 0.008} & \citetalias{rosenthal2021} \\
  Radius ($R_\odot$)& \multicolumn{2}{c|}{0.375 $\pm$ 0.007} & \citetalias{rosenthal2021} \\
  Effective temperature (K) & \multicolumn{2}{c|}{3742 $\pm$ 30} &  \citetalias{rosenthal2021} \\
  $H$ magnitude  & \multicolumn{2}{c|}{4.476 $\pm$ 0.2} & \citetalias{cutri2003} \\
  Systemic velocity [\kms] & \multicolumn{2}{c|}{11.73 $\pm$ 0.0001} &  \citetalias{fouque2018} \\
  Limb Darkening (Quadratic) & \multicolumn{2}{c|}{0.0156, 0.313} &  \citetalias{claret2011} \\
   \hline
 \textbf{Planet parameters} &  & & \\
  & HD 189733 b & Synthetic planet & Reference \\ 
  \hline
  Transit depth (\%) & 2.2 $\pm$ 0.1 & 2.2& \citetalias{addison2019}  \\
  Radius ($R_J$) & 1.142 $\pm$ 0.04 & 0.55 & --   \\
  Mass ($M_J$) & 1.13 $\pm$ 0.05  & 0.572& --  \\
  g (m.s$^{-2}$) & 22.45 $\pm$ 1.5 & 49.18& -- \\
  Orbital period (d)  & 2.218577 $\pm$ 0.000001& 2.218577& -- \\
  Mid transit time (BJD TBD) & 2458334.990899 $\pm$ 0.0007 & 2459130.8962180 & \citetalias{addison2019} \\
  Inclination (deg) & 85.7 $\pm$ 0.1 & 90.0& -- \\
  Eccentricity & 0.0  & 0.0& -- \\
  Equilibrium temperature (K) & 1209 $ \pm$ 11 & 1209& - \\
  Orbital semi-amplitude (km.s$^{-1}$) & 151.2 $\pm$ 4.5 & 120.0& \citetalias{addison2019}  \\
  Transit duration (h) & 1.84 $\pm$ 0.04  & 1.84& --  \\
  \hline
\end{tabular}
     \begin{tablenotes}
     \footnotesize
     \item[$\dagger$] To gain some space in the table, we use aliases for the references. \citetalias{rosenthal2021}, \citetalias{cutri2003}, \citetalias{fouque2018}, \citetalias{claret2011} and \citetalias{addison2019} stand respectively for \citet{rosenthal2021}, \citet{cutri2003}, \citet{fouque2018}, \citet{claret2011} and \citet{addison2019}.
    \end{tablenotes}
    \end{threeparttable}   
    \caption{Physical parameters for Gl\,15A, HD189733 b and for the simulated hot Jupiter used in the study. When taken from the literature, the reference of each parameter is indicated in the right-hand column$^{\dagger}$.}
\label{tab:parameters}
\end{table*}

\section{Data generation and analysis}
\label{sec:Data}

The generation of the synthetic spectra and their reduction are extensively described in paper~I. They are very shortly reminded here. 

\subsection{Creation and reduction of synthetic data}

\begin{figure}
    \centering
    \includegraphics[width=\linewidth]{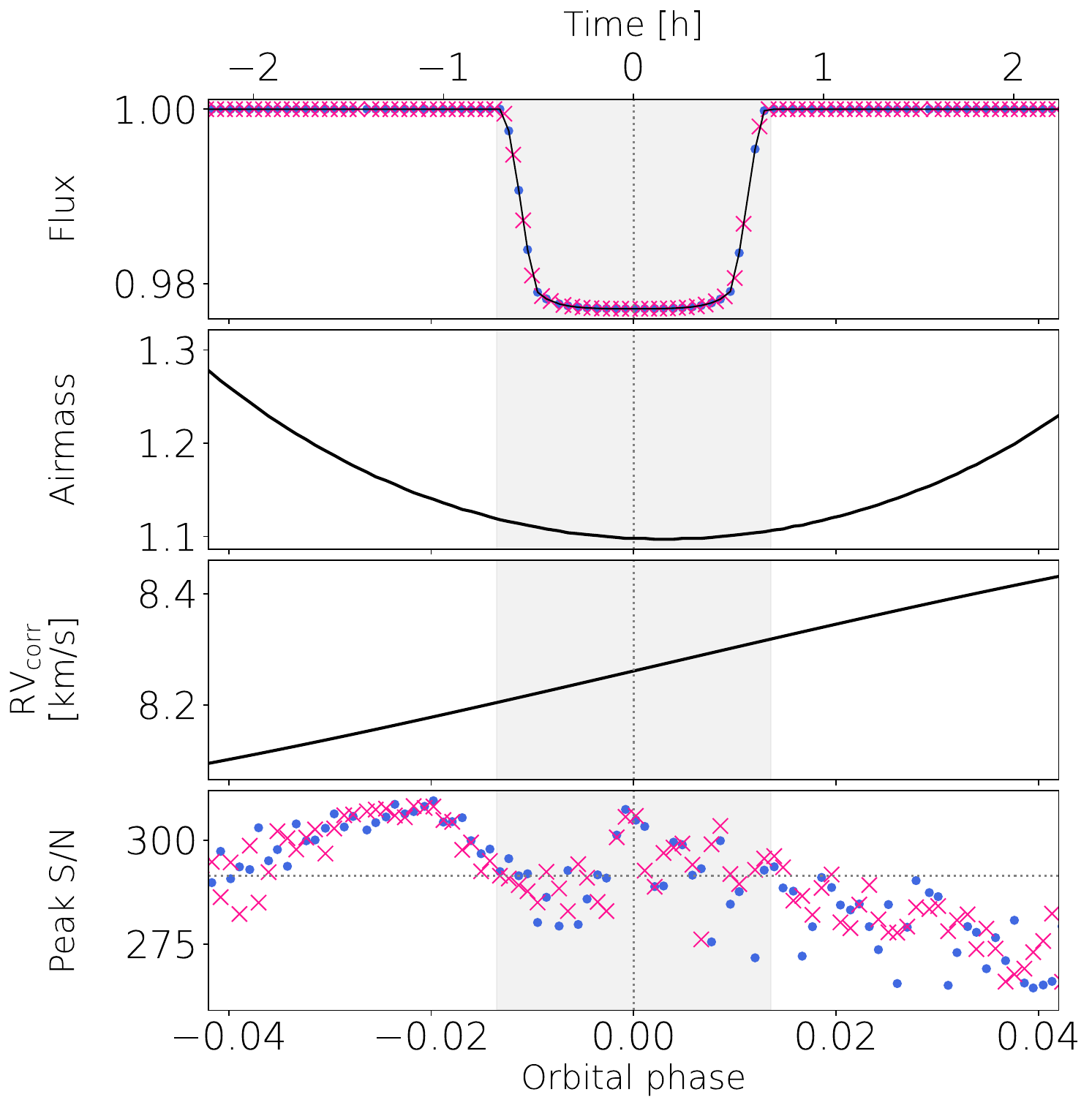}
    \caption{Variations of photometric flux (top panel), airmass (panel~2), Geocentric-to-stellar rest frame RV correction (panel~3) and peak SNR per velocity bin during the two simulated transits of the \hdb\ analog. On panels~1 and~4, the two different transits are respectively shown as blue dots and pink crosses. The vertical gray band indicates the primary transit of the simulated planet. The horizontal gray dashed line on the bottom panel indicates the average value of the peak S/N for the observed spectra.}
    \label{fig:transit_info}
\end{figure}

We simulate the observations of a planetary transit with a near-infrared (nIR) high-resolution spectrograph. This is done by injecting a synthetic planet atmosphere spectrum into a sequence of 192 spectra of the bright M dwarf Gl 15A collected during 5 hours with SPIRou in October 2020 and divided in two sets (see Table\,\ref{tab:parameters} and Figure~\ref{fig:transit_info}). \GLA is chosen both because we have many spectra of it with SPIRou and because it is a well-studied star in radial velocity. If its system contains  a short-period Earth-like planet, it is shown that no Jovian planet orbits this star for periods of less than 10 years (see for example Figure 2 of \citet{Pinamonti2022}.The observations sample the [0.9,2.5] $\mu$m wavelength range in 49 diffraction orders with a typical pixel size of $2.28$ km.s$^{-1}$ and a spectral resolution of $\sim 70 000$. Data are reduced through the APERO pipeline \citep{cook2022} that calibrates the data in wavelength and applies state-of-the art telluric correction.

The synthetic planet is based on the classical hot Jupiter HD189733 b \citep{bouchy2005} injected on a circular orbit and we decided to conserve four planetary and transit parameters to obtain data with consistent expected level of detection: (i)~the transit depth, (ii)~the transit duration,  (iii)~the ratio between the stellar radius and the atmospheric scale height and (iv)~the atmospheric temperature. The injected planet spectra are all shifted by $30$ \kms\ so that stellar molecular features and planet atmosphere absorption lines are not mixed.

Once the planet is injected, we remove the stellar spectra and remaining telluric contaminations by dividing each observed spectrum within each order by a median spectrum. This step is performed successively in the Earth rest and stellar rest frames, and an additional high-pass filter is applied to the residual spectra, in order to correct for low-frequency variations in the continuum. Outliers are flagged and masked using a sigma-clipping procedure, and the residual time-varying telluric flux is corrected with an airmass detrending in the log-flux space. We then apply a principal component analysis (PCA) to get rid of the remaining correlated noise. An auto-encoder can be applied instead, although it is not yet mature for parameter retrieval and is limited to detection of molecular species as we cannot reproduce its effect efficiently on the models. Diffraction orders~57 to~54 (i.e. $\sim$1\,300 to $\sim$1\,500\,nm) and 42 to 40 (i.e., $\sim$1\,800 to $\sim$2\,000\,nm), located within nIR water absorption bands, are discarded.


\subsection{Uncovering the planetary signature}

Once the reduced data have gone through the PCA or auto-encoder step, the planetary signal is still largely buried under the noise. We either perform a template matching method between theoretical models and the reduced data, or a statistical exploration of the parameter space through nested sampling using the python module \texttt{pymultinest} \citep{Buchner2014,Feroz2008,Feroz2009,Feroz2019}. The models are created with petitRADTRANS \citep{Molliere2019} which provides the planerary radius as a function of wavelength. They are next trasnformed into an absorption by calculating the ratio of planetary to stellar radius squared, the so-called transit depth.
The correlation function (CCF) calculated for different planet velocimetric semi-amplitude (\kp) and systemic Doppler shift (\vsys) writes: 

\begin{equation}
    CCF = \sum_i \left. \dfrac{d_i m_i}{\sigma_i^2} \text{,} \right.
\end{equation}
where $m_i$, $d_i$ and $\sigma_i$ are respectively the flux in the model spectrum, the observed flux and the flux uncertainty at pixel $i$ (corresponding to time $t$ and wavelength $\lambda$: $d_i = d(t,\lambda)$.). This function is calculated and summed for every SPIRou order. More precisely:

\begin{equation}
    \sigma_i^2 = \sigma^2 (t,\lambda) = \dfrac{\sum_t \left( d(t,\lambda)-\overline{d(\lambda)}\right)^2}{N_\mathrm{spectra}}\dfrac{\overline{\mathrm{SNR}}}{\mathrm{SNR}(t)}
\end{equation}
where the bar denotes a time average and $N_\mathrm{spectra}$ is the number of spectra. The barred SNR values are calculated for each order. 	In order to convert correlation value to significance of detection, we perform as is frequently done in literature, i.e.,  divide by the standard deviation of the correlation map away from the planetary signal. 

The nested sampling relies on the calculation of a likelihood $L$, defined following the frameworks of \citet{Brogi2019} and \citet{Gibson2020}:

\begin{equation}
    \mathcal{L} = \prod_{i=0}^{N} \dfrac{1}{\sqrt{2 \pi \sigma_i}} \mathrm{exp} \left\{ -\dfrac{[m_i-a d_i]^2}{b \sigma_i^2} \right\} \text{,}
\end{equation}
where $a$ and $b$ are scaling factors to account for incorrect modelling. a is set to 1 in this paper, and b is optimized globally as in \citet{Gibson2020}. 

In order to account for the fact that the observed planet atmosphere spectrum is affected by the data reduction procedure, we degrade the model before comparing it to the data, following the procedure detailed in \citet{Gibson2022}. First, we create the projector $\mathrm{P}$  on the vector space defined by our subset of PCA eigenvectors obtained in the data analysis. At each iteration of the nested sampling process, we then compute a sequence of model spectra, called $\mathrm{M}$, matching the wavelength and time grids of the observations, and Doppler shifted are the values of \kp and \vsys. 
We finally subtract the projection by $\mathrm{P}$ of $\mathrm{M}$  . Our final, degraded sequence of theoretical model
$\mathrm{M}'$ is:

\begin{equation}
    M' = \exp \left(\log M - P \log M \right) \text{.}
\end{equation}
As we show in our companion paper, this step is crucial not to bias the retrieved planet parameters.

Finally, as explained in our companion paper,  we have the possibility to include a proxy for planetary rotation and winds. We simply convolve our 1D atmospheric models by a rotation kernel that considers the latitudinal speed variation due to rotation. This kernel can be modulated to take into account any latitudinal wind shape, such as superrotation. We expressed this kernel as two convolution products so that it is very efficient numerically speaking and allows one to retrieve the planet rotation rate in a parameter space exploration algorithm.

\section{Application to simulated data}\label{sec:Results}
\label{sec:results}

\subsection{Uncertainties in mass and radius}

Our first test was to keep a simple, isothermal model containing only water as in paper I but to change the radius $R$ and gravitational acceleration $g$ (proxy for mass $M$ here as $g \propto M/R^2$) of the planet in the retrieval compared to the injected planet.  We tested three different cases: in Model A, gravity and radius are treated as free parameters in the retrieval. Model B imposes a wrong gravity and a good radius and we look at the effect on the temperature and water composition. Model C imposes a wrong gravity but allows only the radius to change. The results  and their comparison with the isothermal model of Paper 1 are provided in Table \ref{tab:massrad}. Through the section, $R$ and $g$ are expressed in planetary units.

\begin{figure}
    \centering
    \includegraphics[width=0.4\textwidth]{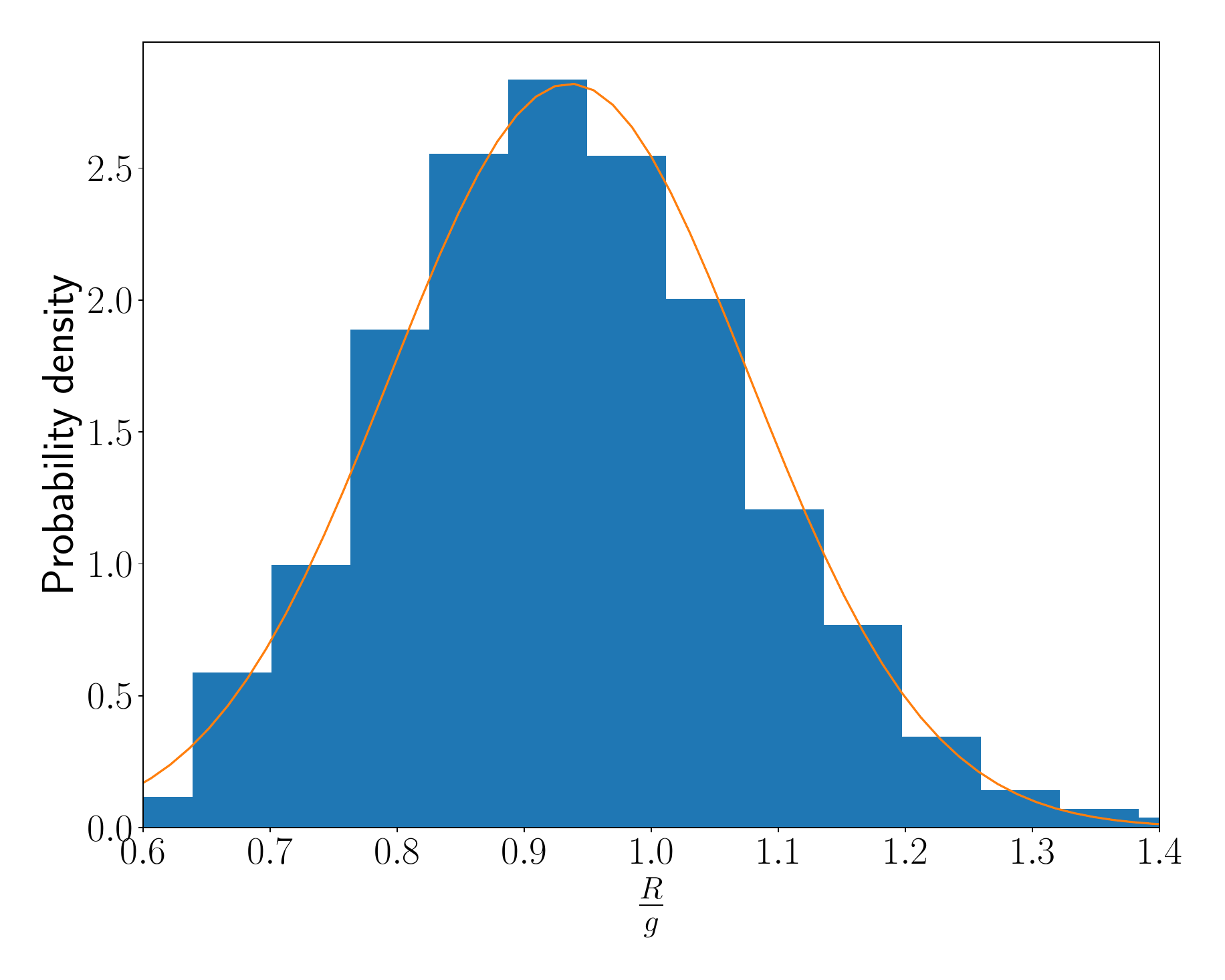}
    \caption{Posterior distribution of $\frac{R}{g}$ from Model A in planetary units. The injected ratio is 1. The orange line is a Gaussian with mean 0.936 and standard deviation 0.14}
  \label{fig:Rg}
\end{figure}

Regarding Model A, gravity and radius are recovered within a few percents of error. There is a large degeneracy between both parameters which translates into a smaller error bar on the normalized $R/g$ ratio: the posterior distribution of $R/g$ shown in Fig. \ref{fig:Rg} shows that the distribution is well matched with a Gaussian of standard deviation $0.14$. The uncertainties on other physical parameters are comparable with paper I and the retrieval is globally comparable with paper I.

When we simply imposed a lower gravity in the nested sampling algorithm compared to the injected model in Model B, the retrieval of composition and temperature gave lower values for mass mixing ratio and temperature. The water mass mixing ratio (MMR) is more affected, with the retrieved value 3 $\sigma$ away from the injected value. However, contrary to the results of paper I, the input parameters are outside of the posterior distribution in the temperature-composition joint posterior(figure not shown): there is an actual bias that was not present in our analysis with correct mass and radius.  

Finally, Model C shows that varying only gravity or radius allows to obtain comparable results with Model 1 for water and temperature, with retrieved $R/g$ close to 1.

\begin{table*}
\centering
\centering
\begin{threeparttable}[Hbt]
\begin{tabular}{ccccccccccccc}
  \hline
 & \multicolumn{3}{c}{\textbf{R (R$_p$)}} & \multicolumn{3}{c}{\textbf{g (g$_p$)}} & \multicolumn{3}{c}{\textbf{$T_\mathrm{eq}$ (K)}} & \multicolumn{3}{c}{\textbf{Water MMR }} \\
 & True & Input & Retrieved & True & Input & Retrieved & True & Input & Retrieved & True & Input & Retrieved \\
 \cmidrule(lr){2-4}\cmidrule(lr){5-7}\cmidrule(lr){8-10}\cmidrule(lr){11-13}
 P I & 1.0 & 1.0 & - & 1.0 & 1.0 & - & 900 & [200,2000] & 1013 $\pm$ 117 & -2.11 & [-8,-1] & -2.49 $ \pm$ 0.41 \\
 M A & 1.0 & [0.6,1.4] & 0.94 $\pm$ 0.17 & 1.0 & [0.6,1.4] & 1.04 $\pm$ 0.18 & 900 & [200,2000] & 1026 $\pm$ 124 & -2.11 & [-8,-1] & -2.57 $\pm$ 0.44 \\
 M B & 1.0 & 1.0 & - & 1.0 & 0.75 & - & 900 & [200,2000] & 967 $\pm$ 119 & -2.11 & [-8,-1] & -3.05 $\pm$ 0.37 \\
 M C & 1.0 & [0.5,1.5] & 0.725 $\pm$ 0.13 & 1.0 & 0.75 & - & 900 & [200,2000] & 1003 $\pm$ 130 & -2.11 & [-8,-1] & -2.61 $\pm$ 0.44 \\
 \hline
\end{tabular}
\end{threeparttable}
\caption{Summary of the retrieved parameters when varying mass and radius. The radius is in true planetary radius (R$_p$), gravity in planetary gravity (g$_p$) and water mass mixing ratio (MMR) in log. For each parameter, the first column is the model true value. The second column represent input values when only one number is provided or uniform prior range when the parameter is included in the retrieval. The third column is the retrieved values with $1 \sigma$ uncertainty. The description of the models is given in the text. P I is paper one, and M x means Model x. }
\label{tab:massrad}
\end{table*}

The results of this section show that, within our high SNR framework, we are sensitive to more than the sole amplitude of molecular lines. Indeed, observationnaly speaking we are sensitive to the variations of the transit depth with wavelength hence the important quantity is:

\begin{equation}
    O \sim \dfrac{H R_\mathrm{p}}{R_\mathrm{s}^2}\text{,}
\end{equation}
where $H$ is the typical scale height of the atmosphere, $R_\mathrm{p}$ the planetary radius and $R_\mathrm{s}$ the stellar radius. For the simple case of an isothermal atmosphere:
\begin{equation}
    H = \dfrac{\mathcal{R}T}{M g} \text{,}
\end{equation}
where $\mathcal{R}$ is the ideal gas constant, $T$ the temperature, $M$ the molecular mass and $g$ the gravitational acceleration. If the amplitude was the only concern, we would find that the mass and radius of the planet would be correlated with temperature as the code tries to match the $H R_\mathrm{p}$ value to that of the injected planet. The fact that temperature and composition uncertainties remains globally insensitive to mass and radius as long as $R/g \approx 1$ shows that we are not only matching amplitudes, but the shape of the lines as well which are affected by temperature and composition only.

This analysis points towards the fact that the uncertainty in mass and radius should be included in the retrieval of atmospheric parameters rather than chosen as constants. It allows to avoid biases and, at least in our simple case, does not degrade the obtained atmospheric properties. For optimization purposes, only one of these quantities can be included in the retrieval, remembering that we are only sensitive to the ratio of radius and gravity. This will be particularly stringent for low-mass, distant exoplanets or planets around very active/young stars where the complicated radial velocity signature might lead to large uncertainties in the mass.

\subsection{Non vertically isothermal models}

\begin{figure*}
    \centering
    \begin{subfigure}[t]{0.48\textwidth}
    \includegraphics[width=\textwidth]{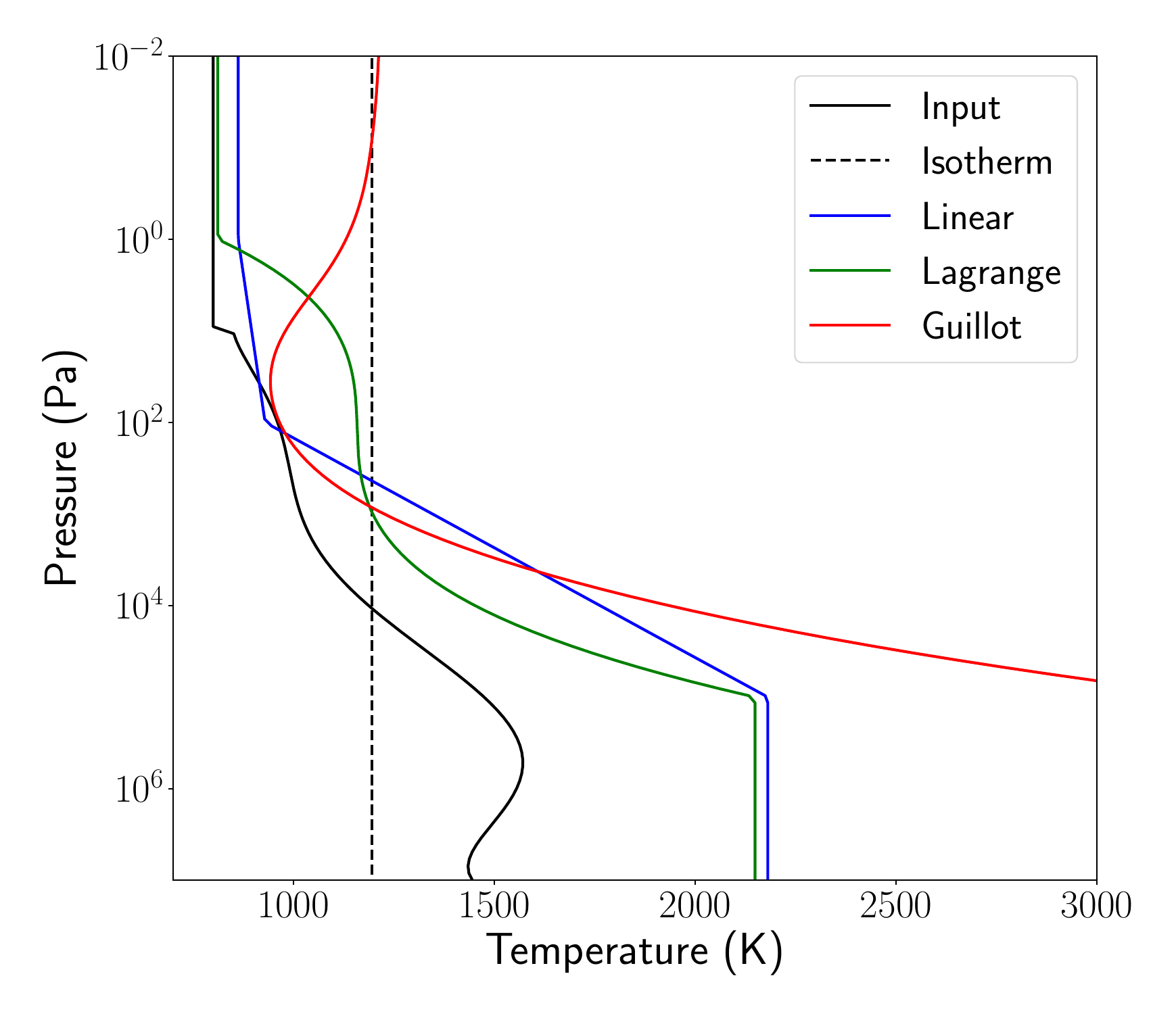}
    \label{fig:temph}
    \end{subfigure}
    \hfill
    \begin{subfigure}[t]{0.48\textwidth}
    \includegraphics[width=\textwidth]{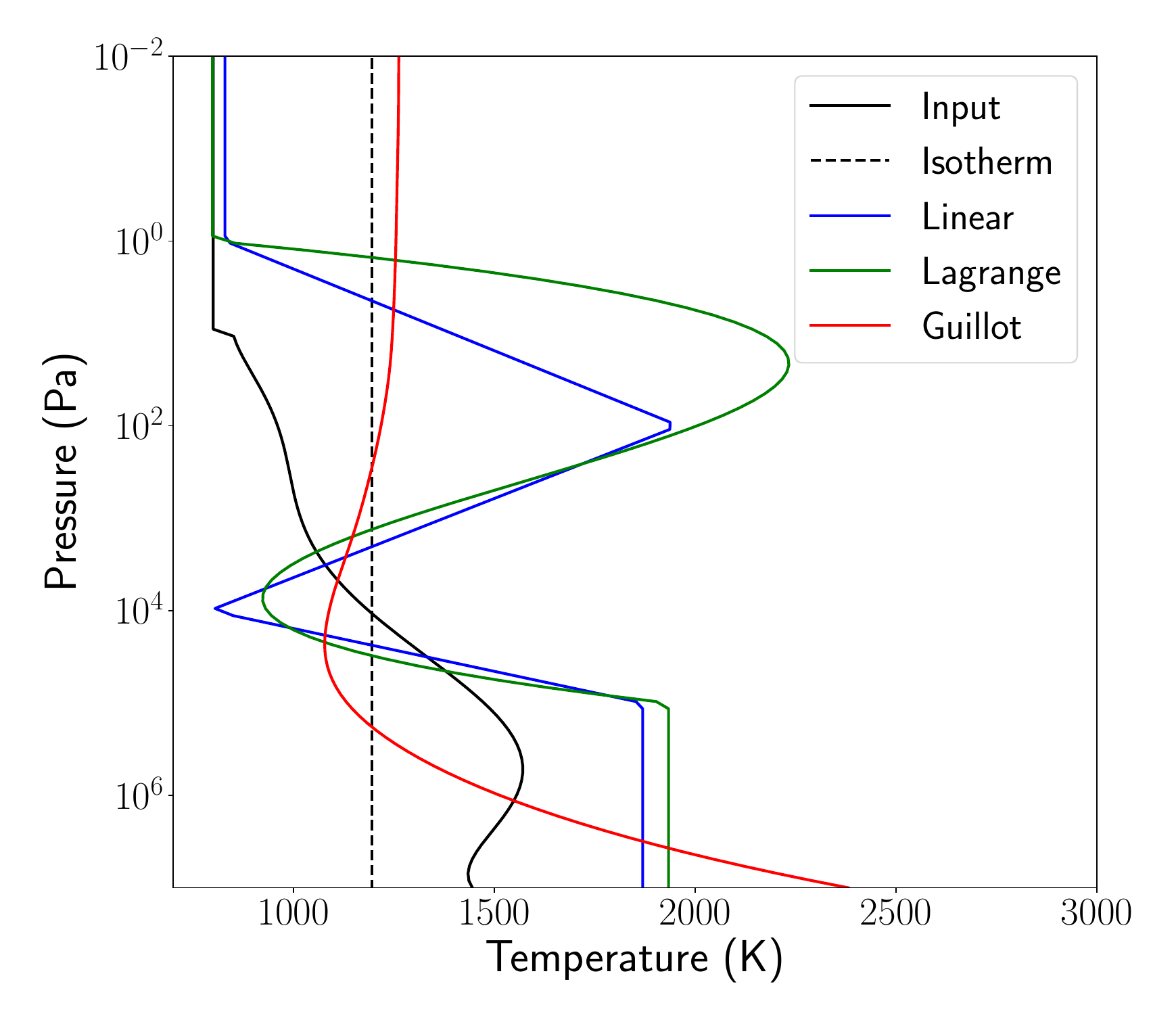}
    \label{fig:templ}
    \end{subfigure}
    \caption{Temperature as a function of pressure for the input profile (black) and the mean of the retrieved profiles with the three possible models discuted in the text (colors). Left: water volume mixing ratio of  $10^{-3}$. Right: water volume mixing ratio of  $10^{-5}$}
  \label{fig:temp}
\end{figure*}

In paper I, we only considered vertically-constant models in temperature and composition. Here, we test whether we are able to retrieve parameters that vary vertically, focusing on non-isothermal profiles. We implement a vertical temperature profile taken from 3D GCM simulations of HD 189733 b \citep{Drummond2018}. The temperature structure is averaged at both limbs and used as an input in the 1D \texttt{petitRADTRANS} modelling (see Fig. \ref{fig:temp}). We created two models with constant water volume mixing ratio (VMR) of $10^{-3}$ and $10^{-5}$ respectively. 

\begin{figure*}
    \centering
    \begin{subfigure}[t]{0.48\textwidth}
    \includegraphics[width=\textwidth]{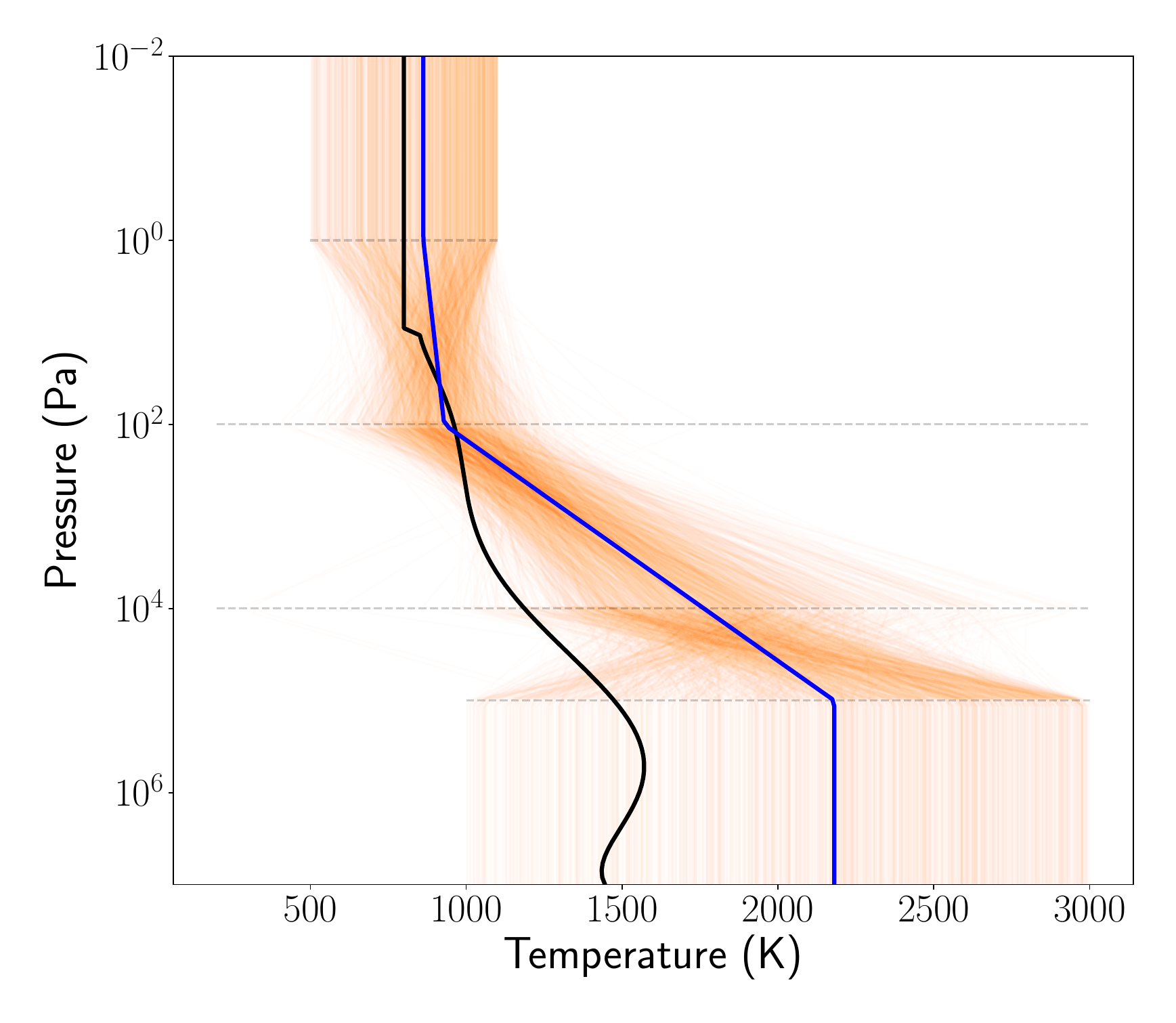}
    \label{fig:tempallh}
    \end{subfigure}
    \hfill
    \begin{subfigure}[t]{0.48\textwidth}
    \includegraphics[width=\textwidth]{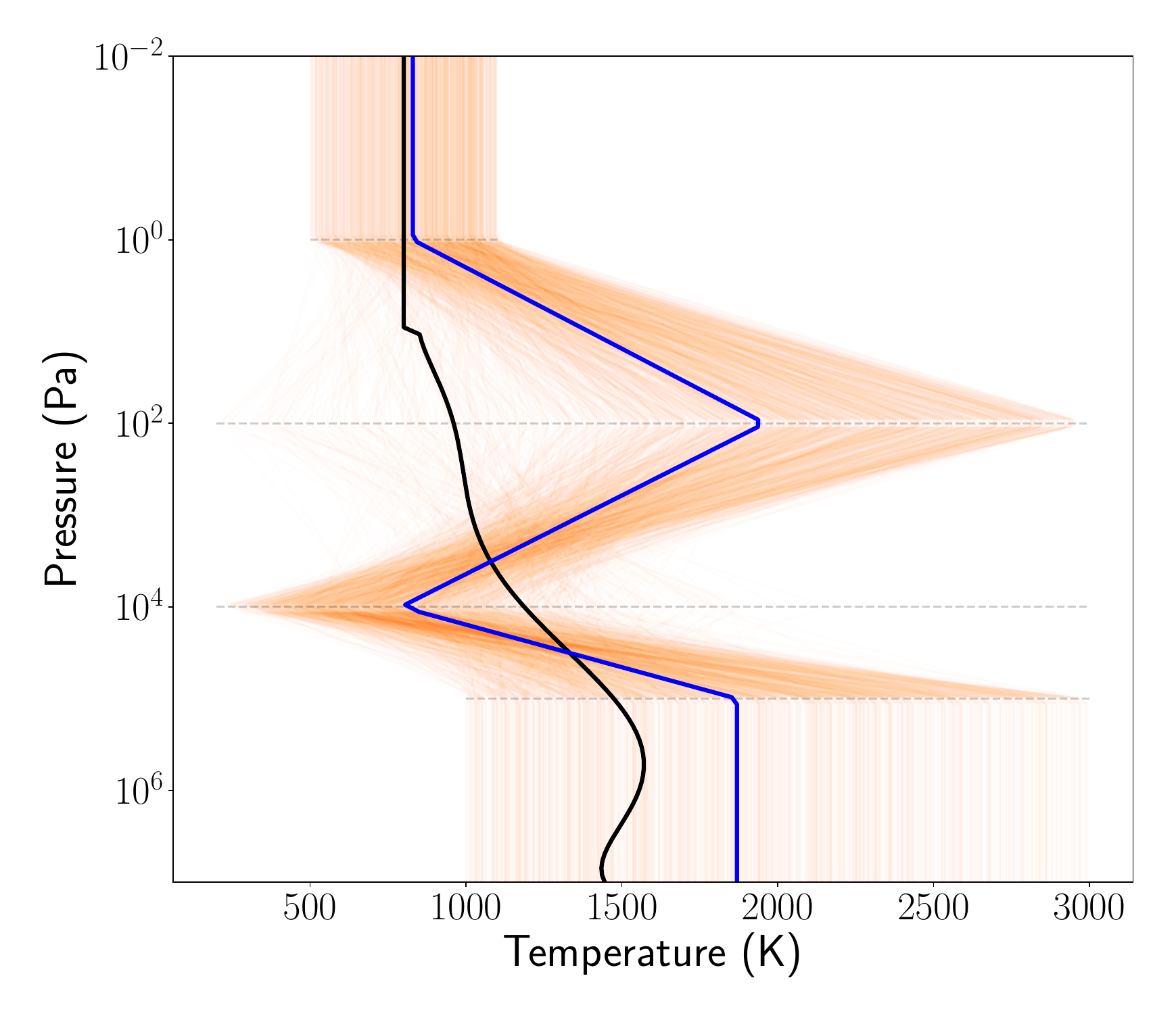}
    \label{fig:tempalll}
    \end{subfigure}
    \caption{Temperature as a function of pressure for the input profile (black) and all of the retrieved profiles from the linearly interpolated 4 points retrieval, with the mean profile in blue. The grey dashed lines represent the uniform prior ranges at the 4 pressures. Left: water volume mixing ratio of  $10^{-3}$. Right: water volume mixing ratio of  $10^{-5}$.}
  \label{fig:temp_all}
\end{figure*}

For our retrieval, we have tested four temperature prescription: in case "Isotherm" we assumed an isothermal profile. Case "Linear" used a 4 points temperature profile, where we retrieve the temperature at 4 different pressures (1 Pa, 100 Pa, $10^{4}$ Pa and 1 bar) and linearly interpolate in log pressure between these points. The temperature at lower pressures than 1 Pa and higher than 1 bar is set as constant. Case "Lagrange" also used the same 4 points prescription but the interpolation was made through a Lagrange polynomial, ensuring a smooth temperature profile. Finally in case "Guillot" , we have used the widely applied two temperature model of \citet{Guillot2010} and retrieved 4 parameters: the internal temperature, the equilibrium temperature, the infrared opacity and the infrared to visible opacity ratio. As we wanted to focus on the temperature structure, we present results where we have fixed the water composition in the retrieval to the model composition. When we let this value as a free parameter, we always retrieved the appropriate water composition but the temperature retrieval is worsen due to expected (and already presented in paper 1) degeneracies. The retrieved temperatures profiles are shown in Figs. \ref{fig:temp} and \ref{fig:temp_all}.

We first notice that the temperature profile is poorly constrained: the standard deviation with pressure can easily reach hundreds of Kelvin, which is much more than the $\approx$ 115K we obtained in the isothermal cases of paper I. The temperature is on average higher than the injected profile, but this mainly arises from the choice of priors which are not centered around the injected profile. In both cases, as seen in Fig. \ref{fig:temp_all}, the deep ($\ge$ 1 bar) and shallow ($\le$ 1 Pa) atmosphere temperatures are just given by the priors: the distribution is close to the uniform prior distribution we chose. However, what we clearly see on Figs. \ref{fig:temp} and \ref{fig:temp_all} is that the retrievals are sensitive to different regions for the two models: the high (low) water concentration model is more sensitive to the higher (deeper) atmosphere. This was expected as the retrieved radius depends mostly on the regions which contribute the most to the water absorption lines, which correspond to pressures where the water column becomes optically thick (optical depth becoming greater than 1). This roughly corresponds to 100~Pa in the dense water model, and $10^4$ in the other. 

Interestingly, although the mean profile is closer to the injected profile around $10^4$ Pa in the low VMR case, the standard deviation of recovered temperature with pressure is always lower in the high VMR case. This arises from the lower SNR in the low VMR case: even at the pressures which contains most of the water information, the amplitude of the planetary signal is too low to permit a precise fit to the data. This is consistent with the isothermal retrieval, whose posterior temperature distribution is well matched by a gaussian with mean and variance 1041$\pm$63 K in the high VMR case, and 1195$\pm$143 K in the low VMR case. 

This test therefore shows that, on average, we are poorly sensitive to the temperature structure and that we are primarily probing pressures where the optical depth of molecular lines reaches 1. This means that we could potentially obtain a better understanding of the temperature profile by combining the information from different molecules: depending on their density and opacity, they will probe different pressure levels. A retrieval with a unique temperature profile for different molecules might therefore be less informative than trying to retrieve the temperature for different species and estimating where they provide most of their signal. 

\subsection{Recovering a multi-species model}

We now consider the case of multiple species, namely H$_2$O, CO, CH$_4$ and NH$_3$. Through this section, unless specified otherwise our synthetic atmosphere always contains these four molecules (in addition to H$_2$ and He which are largely dominant) with isothermal profile and we only vary the VMRs of each individual species.

\begin{figure*}
    \centering
    \begin{subfigure}[t]{0.7\textwidth}
    \includegraphics[width=\textwidth]{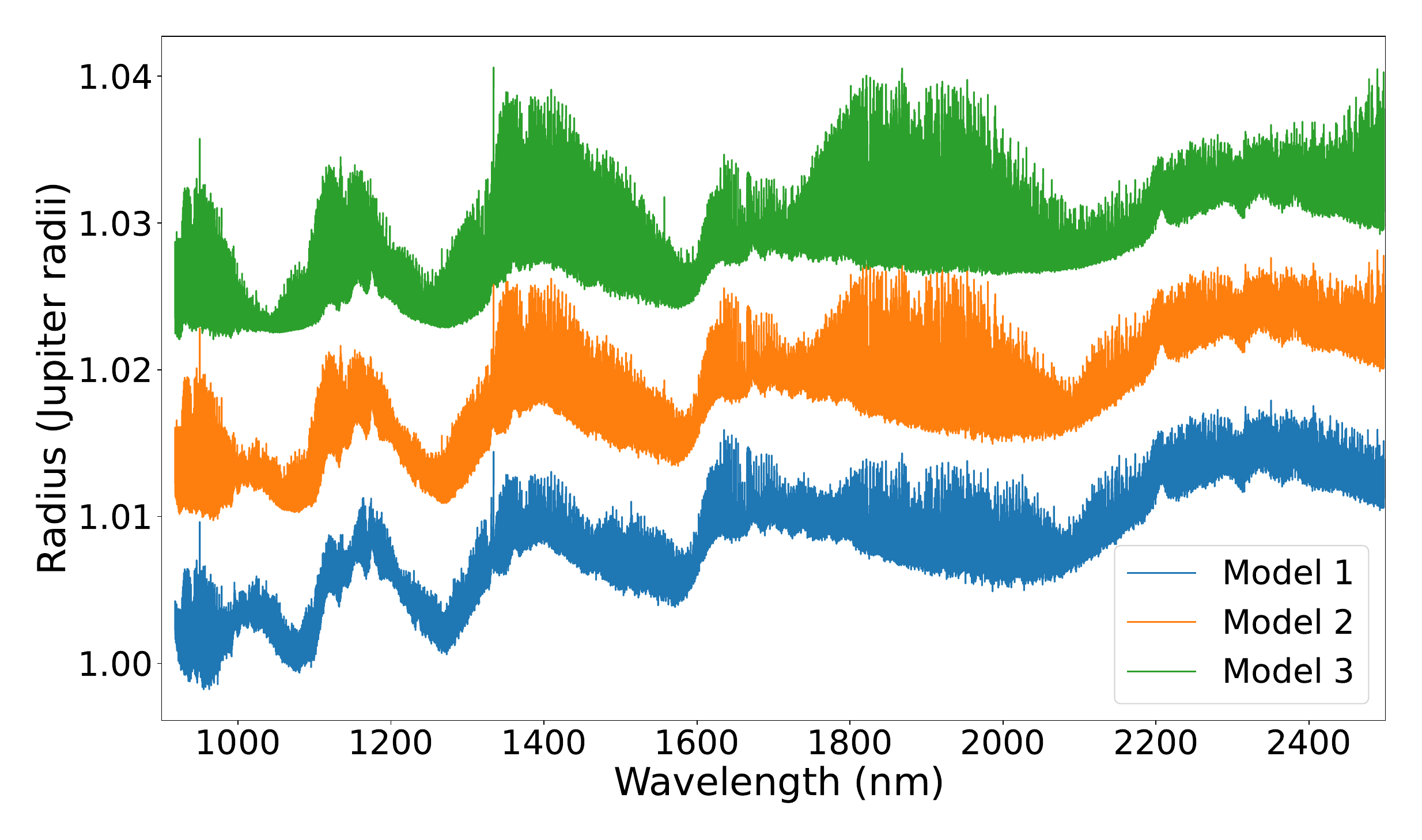}
    \label{fig:H2O-vs-multi}
    \end{subfigure}
    \\
    \begin{subfigure}[t]{0.48\textwidth}
    \includegraphics[width=\textwidth]{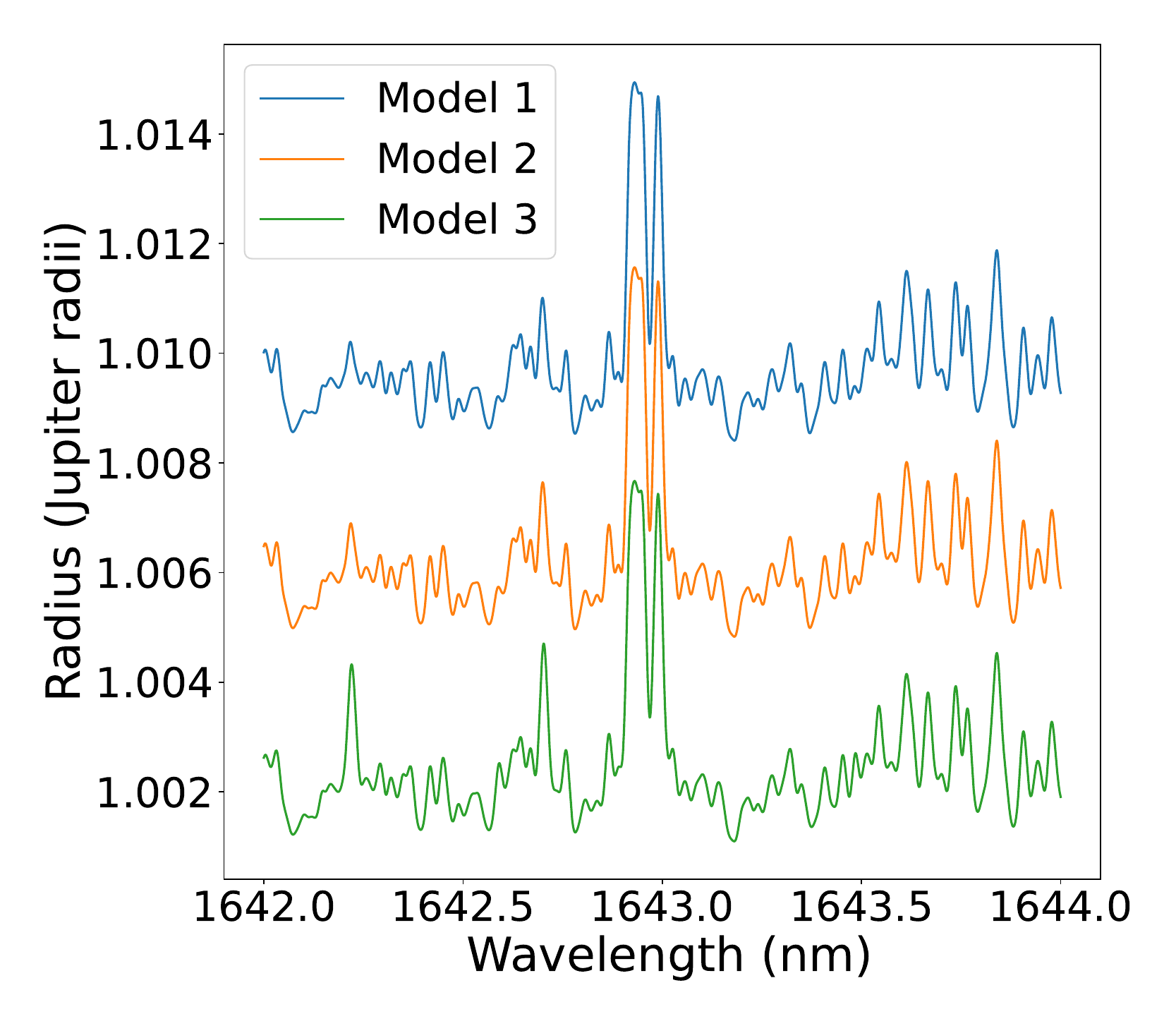}
    \label{fig:H2O-vs-multi-zoom1}
    \end{subfigure}
        \begin{subfigure}[t]{0.48\textwidth}
    \includegraphics[width=\textwidth]{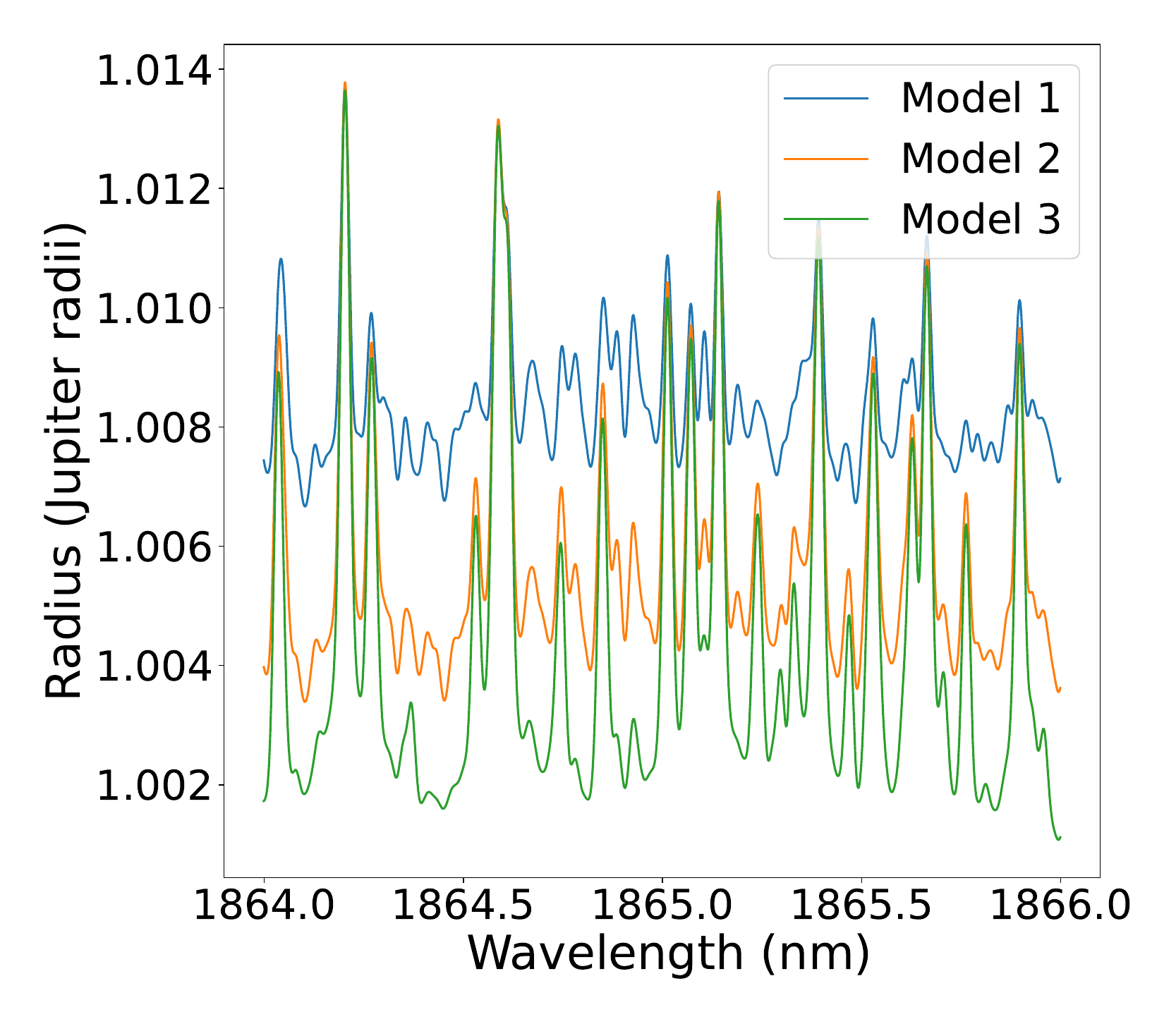}
    \label{fig:H2O-vs-multi-zoom2}
    \end{subfigure}
    \caption{Top: transit radius as a function of wavelength over the SPIRou domain for the three models of Table \ref{tab:multi}. The models have been shifted for visual comparison. Bottom: zoom on two different wavelength ranges. The models have not been shifted in the zooms.  } 
  \label{fig:H2O-multi-tot}
\end{figure*}

We created three synthetic transit sequences with three injected models labelled 1, 2 3. Model 1 used the MMRs reported in the table 4 of \citet{Giacobbe2021}, Model 2 kept the same MMR for water but the other molecules were a factor of 10 lower and Model 3 a factor of 100 lower. The characteristics of the three models are  summarised in Table \ref{tab:multi} and plotted on Fig. \ref{fig:H2O-multi-tot}, where we show the whole models and two zooms: one where water has low-amplitude absorption lines (around $1640$ nm) and one where water absorption is dominant (around $1860$ nm).

\begin{table}
\centering
\begin{threeparttable}
\begin{tabular}{cccc}
  \hline
  & \textbf{Model 1} & \textbf{Model 2} & \textbf{Model 3}\\
  \hline
  Temperature & 900 & 900 & 900\\
  log$_{10}$(H$_2$O) & -3.05 & -3.05 & -3.05\\
  log$_{10}$(CO)  & -1.8 & -2.8 & -3.8\\
  log$_{10}$(NH$_3$)  & -3.0 & -4.0 & -5.0\\
  log$_{10}$(CH$_4$)  & -1.5 & -2.5 & -3.5\\
  \hline
\end{tabular}
\end{threeparttable}   
\caption{Physical parameters for the multi species models included used in the nested sampling retrieval. The temperature is in Kelvin and the abundances in mass mixing ratios, to be easily comparable with the posterior figures.}
\label{tab:multi}
\end{table}

In Appendix \ref{app:multi}, we show the cross-correlation maps for the three synthetic models.  When we correlated the synthetic data with the injected models in Fig. \ref{fig:corr_tot} , we recovered $\approx 5\sigma $ detection in all cases, with Model 3 having the highest SNR and ratio between maximum positive and minimum negative value of correlation. This is not surprising as wee see on Fig. \ref{fig:H2O-multi-tot} that the amplitude of the lines is higher for this model, where the lower level of other species impacts less the global shape of the spectrum. 

We then correlated the synthetic data with models containing only one of the species. This is usually done in the literature for atmospheric characterisation to validate a detection of an individual species even if the atmosphere contains other constituents. When using all of the orders, CH$_4$ was detected (significance larger than 3$\sigma$) for all 3 models as shown in Fig.\ref{fig:corr_CH4-3} for Model 3. NH$_3$ on the other hand is detected in Model 1, only marginally detected (detection around 2 $\sigma$) in Model 2 as shown in Fig. \ref{fig:corr_NH3-12} and not detected in Model 3. Water is not detected in Model 1, detected in Model 2 as shown in Fig. \ref{fig:corr_H2O-23} and detected over 4 $\sigma$ in Model 3 . Finally, we never detected CO as we explain in the next paragraph. These rather poor results led us to consider selecting orders, as we detail in the next section. 

Importantly, NH$_3$ and H$_2$O in Model 1 and 2 were not robustly detected although, when we performed injection-recovery test with only these molecules at the same VMR they were easily detected (larger than 4 $\sigma$). This shows that the non detection of a given individual species does not systematically mean that it is absent from the atmosphere but can simply reflect a mismatch between a complex observed atmosphere spectrum and a too simplistic single-species model.

\begin{table}
\centering
\centering
\begin{threeparttable}[Hbt]
\begin{tabular}{cccc}
  \hline
 & {\textbf{H$_2$O}} & {\textbf{NH$_3$}} & {\textbf{CH$_4$ }} \\
Model 1 & No & \yes{Yes} & \yes{Yes} \\
Model 2 &\yes{Yes} & Marginal & \yes{Yes} \\
Model 3 & \yes{Yes} & No & \yes{Yes} \\
  \hline
\end{tabular}
\end{threeparttable}
\caption{Detection of individual species when correlating single-component models with the synthetic data detailed in Table \ref{tab:multi}. A detection means a SNR superior to 3, whereas a marginal detection is between 2 and 3. }
\label{tab:multi_single}
\end{table}

For CO, we realised that the issue came from the stellar CO which prevents the detection of planetary CO. When we divide by the mean stellar spectra in the data reduction process we affect the planetary lines and hamper the detection. However, we also tested that the presence of CO in the synthetic data only marginally affected the retrieval of other species, due to its limited wavelength range of absorption and well separated absorption lines. We will therefore not consider CO in the rest of this section although it is in the models. 


We then tried to retrieve the parameters with our nested sampling algorithm. We show the resulting posterior distributions in Appendix \ref{app:marg}. Several things can be noted:

\begin{itemize}
    \item The H$_2$O abundance is poorly constrained in the two first models. This is not surprising as we exclude orders where tellurics are dominant, hence where water has its major impact on the spectrum. It was not an issue in the single species model but becomes problematic when other species are considered with the same VMR as water and the absorption lines of water are reduced in amplitude. In Model 3, we recover results comparable to those of paper I.
    \item The temperature is always over-estimated and leads to large degeneracies with composition. When we fix the temperature, as shown in Fig. \ref{fig:nested_multi_reduced_fixT} compared to Fig. \ref{fig:nested_multi_reduced}, the retrieval of other parameters is largely improved. 
    \item  There is some degeneracy between H$_2$O, CH$_4$ and NH$_3$ although we expect HRS to distinguish between molecular lines of different species. This rather counter-intuitive correlation is easily explained: for a given quantity of, say, H$_2$O, increasing the quantity of CH$_4$ or NH$_3$ decreases the line depth by increasing the mean radius of the planet as seen in Fig. \ref{fig:H2O-multi-tot}. The algorithm thus does not differentiate properly between low quantity of H$_2$O and CH$_4$/NH$_3$ or high quantities of all of them. 
     \item NH$_3$ is not recovered in Model 3 and only an upper limit for its content is obtained. However, we serendipitously removed a few orders where water has the highest signal (around the bands of water at 1.4 and 1.8 microns) and retrieved ammonia in Fig. \ref{fig:nested_multi_reduced100_NH3}. This retrieval shows two peaks : one at low temperature (close to the injected 900K) where  NH$_3$ is poorly recovered but water and methane are, and one at much larger temperatures (few thousands kelvins) where the fit of the NH$_3$ composition is tightly centered around its injected composition, but at the cost of losing the detection of methane and having a degeneracy between water and temperature to ensure a constant line depth for ammonia. We understand this as the fact that the first peak has a maximum of likelihood from the fit of the fewer remaining water lines, whereas the second peak fits perfectly ammonia and provides a secondary maximum. We have verified that, when removing these orders in a pure water model, we do not obtain this second peak and simply recover water at the injected VMR. It is not clear how this result would translate to real planetary observation and whether we could potentially detect ammonia in secondary maxima, but it further confirms the degeneracies between composition and temperature in the amplitude of the lines and that care must be taken from the use of HRS only with simple priors.  
    \item In all cases, the typical error  bar of log-VMR is 2 dex. It shows that an atmospheric retrieval with transmission spectroscopy is powerful to identify species but does not give a precise value of the composition (and temperature), except for much higher SNRs as in \citet{Line2021} or by coupling with LRS. 
\end{itemize}

If we now use the nested sampling process to retrieve individual species from the four-species synthetic data, we confirm the results obtained with the cross-correlation method. If one species is dominating, we are able to recover it with the nested sampling algorithm but if the three have comparable VMRs, they are not always recovered individually. Additionally, even if one species is dominant, we often recover too low an abundance compared to the injected value. This is to be expected: as we see on Fig.  \ref{fig:H2O-multi-tot}, the depth of the absorption lines is reduced by the presence of other species. This translates into a lower recovered value of the VMR compared to the injected one, which is not an error of the algorithm but rather comes from too simple an assumption (that the multi-component model is equivalent to a combination of single-component models). Hence, low VMRs of given species in planetary atmospheres can simply arise from an erroneous chemical composition. 

Finally, we also performed a test in which we aimed to retrieve
species that were not included in the model. We used the retrieval with H$_2$O, CH$_4$ and NH$_3$ with an atmosphere model containing only H$_2$O. The \texttt{multinest} algorithm converged towards a low ($\le 10^{-5}$ ) but non zero composition in CH$_4$ and NH$_3$, because of the degeneracies we already mentioned. It therefore shows that the best fit might not lead to a real individual detection and care must be taken when analyzing only posterior data.

\subsection{Order selection}
As we could not always detect molecules individually when considering all orders, we tried to define a merit function that would select or weight the wavelength range for each molecule.  Two methods were tested:
(i)~we created a model with VMR = 10$^{-4}$ for H$_2$O, NH$_3$ and CH$_4$ and calculated the pearson correlation coefficient with the single species model. For each molecule, we then only selected the orders where this correlation was larger than 0.5. (ii)~We calculated the autocorrelation of the spectra of each individual species order by order and used it as weights in the CCF. 

For CH$_4$, the second method slightly improved the detection but only marginally. For water and NH$_3$, there was no reliable improvement by using either of the two methods. The second method works best for water in Model 2 but the first method is best in Model 1, whereas it is the opposite for NH$_3$, and in all cases the improvement is only marginal. We therefore could not rely on these methods to improve our detection limits in the general cases.

However, as we mentioned in the previous section, removing a few orders dominated by water helped detecting and constraining the NH$_3$ composition with the nested sampling algorithm in Model 3. This shows that, although for individual molecules we did not find reliable ways to improve the significance of correlation, order selection can improve the retrieval for multiple species models. This is to be kept in mind when trying to rule out the presence of certain molecules from Bayesian exploration of the parameter space.

Finally, it is interesting to note that our preferred orders for water detection (following method (i) or taking the 15 best orders of methods (ii)) are very different from those of \citet{Giacobbe2021}. We actually don't recover water in Model 1 or 2, and only marginally in Model 3 with their wavelength domain. This points out toward either a difference in our two analyses, a much larger signal as they combine 5 transits of high atmospheric signature or that HD 209458 b has a much higher water volume mixing ratio than CH$_4$ and NH$_3$.

\section{Discussion}\label{sec:disc}
\label{sec:discussion}

\subsection{Combining transits}

Since addressing all sources of uncertainties and degeneracies is out of reach, we have focused on a few cases but have not mentioned the impact of stacking transits on the retrieval. Obviously, adding up many transits helps in identifying the atmospheric absorption by increasing the SNR as long as there is no (or low) variability in the planetary signature. The combination of several transits will be discussed in detail in forthcoming papers of the ATMOSPHERIX consortium with real data (Masson et al., in prep, Hood et al., in prep.).

\subsection{Improving the detection}

Throughout the two first articles of the ATMOSPHERIX consortium, we have focused on optimising the data reduction process. Further improvement of the data analysis framework will be required to characterise the atmospheres of the most challenging targets of the ATMOSPHERIX sample, either because of their low-amplitude atmosphere signals or due to the host star being too faint and/or active. The community is devoting substantial efforts to enhance the significance of molecular detection and get as much information as possible from the data. The use of an autoencoder, introduced in paper I, is one of such example. Among the possible improvements, we want to mention the works of \citet{Meech2022} and \citet{Ramussen2022}. Both teams use Gaussian processes to perform a spectrum retrieval and improve the data reduction process.


Other technics have been presented in the literature although they have not yet been applied as systematically as template matching. We notably think about tomography \citep{Watson2019} which is an interesting prospect to characterize exoplanets. If we were able to retrieve a mean line profile, Doppler imaging techniques inspired from stellar studies (e.g., \citet{Vogt1987}) could also be used to study the multi-dimensional structure of planets. This prospect is particularly interesting in the visible, where the lines have higher SNR, in emission spectroscopy and in the fortchoming era of 30+ meter telescopes. 

Globally, the use of HRS to characterize exoplanet atmospheres is less than 15 years old, and there are still lots of possibilities to improve the techniques. Such improvements might mitigate  the conclusions of this paper as the detection level will be increased. We still expect degeneracies to be present and important in the process as we exposed them here and we advocate that there is a lack of studies focusing on the inherent degeneracies and limitations of the method thus far.

\section{Conclusion}\label{sec:conclusion}
\label{sec:conclu}

In this paper, we have extended on the work of paper I, presenting our data analysis pipeline, by stuying different sources of uncertainty and degeneracy inherent to our analysis. We have shown that we are able to retrieve the correct model but that numerous degeneracies can drastically increase the error bars. We have focused on three issues: inaccuracies in the mass and radius, non-vertically isothermal profiles and the retrieval of multiple species. The conclusions of our tests are as follows:

\begin{itemize}
\item   The mass and radius of the planet should be included in the retrieval if they are uncertain, as this leads to a more reliable atmospheric retrieval. 
\item The vertical temperature distribution of the planet's atmosphere is not easily retrieved as we are mostly sensitive to pressures where the optical depth approaches to 1. However, this also means that different molecules will be sensitive to different pressure levels which might allow one to probe the atmosphere at different depths, and to reconstruct a global temperature profile by combining information of different molecules at different pressures.
\item Models with multiple species introduce several degeneracies which can lead to erroneous conclusions: one can identify molecules that are not present or estimate inaccurately their mixing ratios.
\item When imposing the temperature, the retrieval is significantly improved.
\item Although transmission spectroscopy is good at detecting molecules, the $1 \sigma$ uncertainty on the volume mixing ratio can reach up to 2 orders of magnitudes for our typical SNR of 200. Stacking many observations or using independent diagnostics such as LRS is necessary to reduce these uncertainties. 
\item We did not find a reliable way to weight or select the SPIRou orders in order to improve the molecular detection for single species. We found however than selecting order can improve the retrieval of one species at the cost of a worse retrieval of another one and temperature in multi-species models.  
\end{itemize}

The combination of this paper and paper I gives an overview of the capacity of our pipeline to analyse SPIRou data  of exoplanet atmospheres through transmission spectroscopy. They will serve as a basis for forthcoming papers of the ATMOSPHERIX consortium on real targets, whose studies are ongoing.

\section*{Acknowledgements}

Based on observations obtained at the Canada-France-Hawaii Telescope (CFHT) which is operated from the summit of Maunakea by the National Research Council of Canada, the Institut National des Sciences de l'Univers of the Centre National de la Recherche Scientifique of France, and the University of Hawaii. The observations at the Canada-France-Hawaii Telescope were performed with care and respect from the summit of Maunakea which is a significant cultural and historic site.
FD thanks the CNRS/INSU Programme National de Planétologie (PNP) and Programme National de Physique Stellaire (PNPS) for funding support. This work was supported by the Action Spécifique Numérique of CNRS/INSU. This work was granted access to the HPC resources of CALMIP supercomputing center under the allocation 2021-P21021. BK acknowledges funding from the European Research Council under the European Union’s Horizon 2020 research and innovation programme (grant agreement no. 865624, GPRV).  JF.D, C.M, I.B. X.B, A.C., X.D.,G.H., F.K acknowledge funding from Agence Nationale pour la Recherche (ANR, project ANR-18-CE31-0019 SPlaSH). AM, BC, BB, SV acknowledge funding from Programme National de Planétologie (PNP) and the Scientific Council of the Paris Observatory. X.D. and A.C. acknoweldge funding by the French National Research Agency in the framework of the Investissements d’Avenir program (ANR-15-IDEX-02), through the funding of the “Origin of Life” project of the Université de Grenoble Alpes.JL acknowledges funding from the European Research Council (ERC) under the European Union’s Horizon 2020 research and innovation programme (grant agreement No. 679030/WHIPLASH), and from the french state: CNES, Programme National de Planétologie (PNP), the ANR (ANR-20-CE49-0009: SOUND). JFD and BK acknowledge funding from the European Research Council (ERC) under the H2020 research \& innovation programme (grant agreement \#740651 NewWorlds).  O.V. acknowledges funding from the Agence National de la Recherche through the ANR project `EXACT' (ANR-21-CE49-0008-01), from the Centre National d'\'{E}tudes Spatiales (CNES), and from the CNRS/INSU Programme National de Plan\'etologie (PNP).
Part of the work has also been carried out in the frame of the National Centre for Competence in Research PlanetS supported by the Swiss National Science Foundation (SNSF). WP acknowledge financial support from the SNSF for project 200021\_200726. PT acknowledges supports by the European Research Council under Grant Agreement ATMO 757858. E.M. acknowledges funding from FAPEMIG under project number APQ-02493-22 and research productivity grant number 309829/2022-4 awarded by the CNPq, Brazil. We wish to thank J. Seidel for her constructive comments and questions.

\section*{Data Availability}

The data are available upon request to the author and the code to analyze them is publicly available.



\bibliographystyle{mnras}
\bibliography{biblio} 



\appendix

\section{Multi species model - all orders}
\label{app:multi}

\begin{figure*}
    \includegraphics[width=0.44\linewidth]{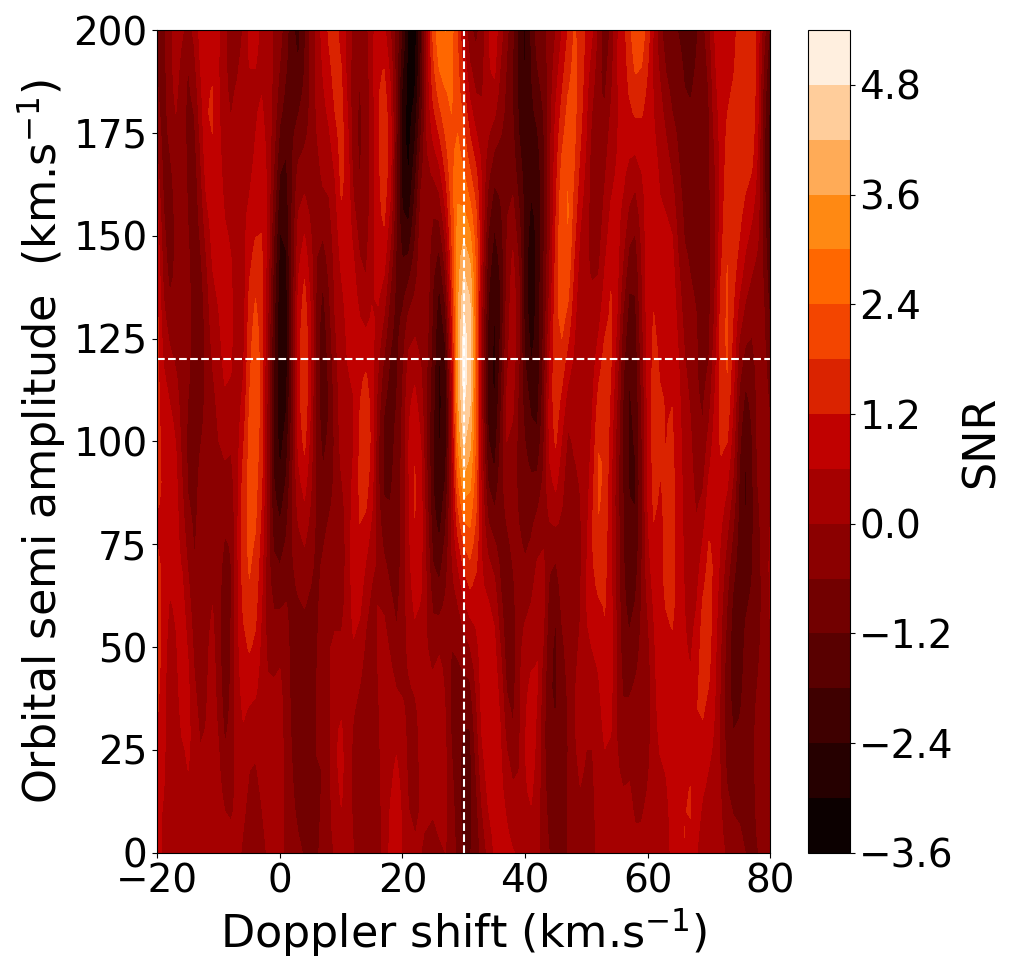}
    \includegraphics[width=0.44\linewidth]{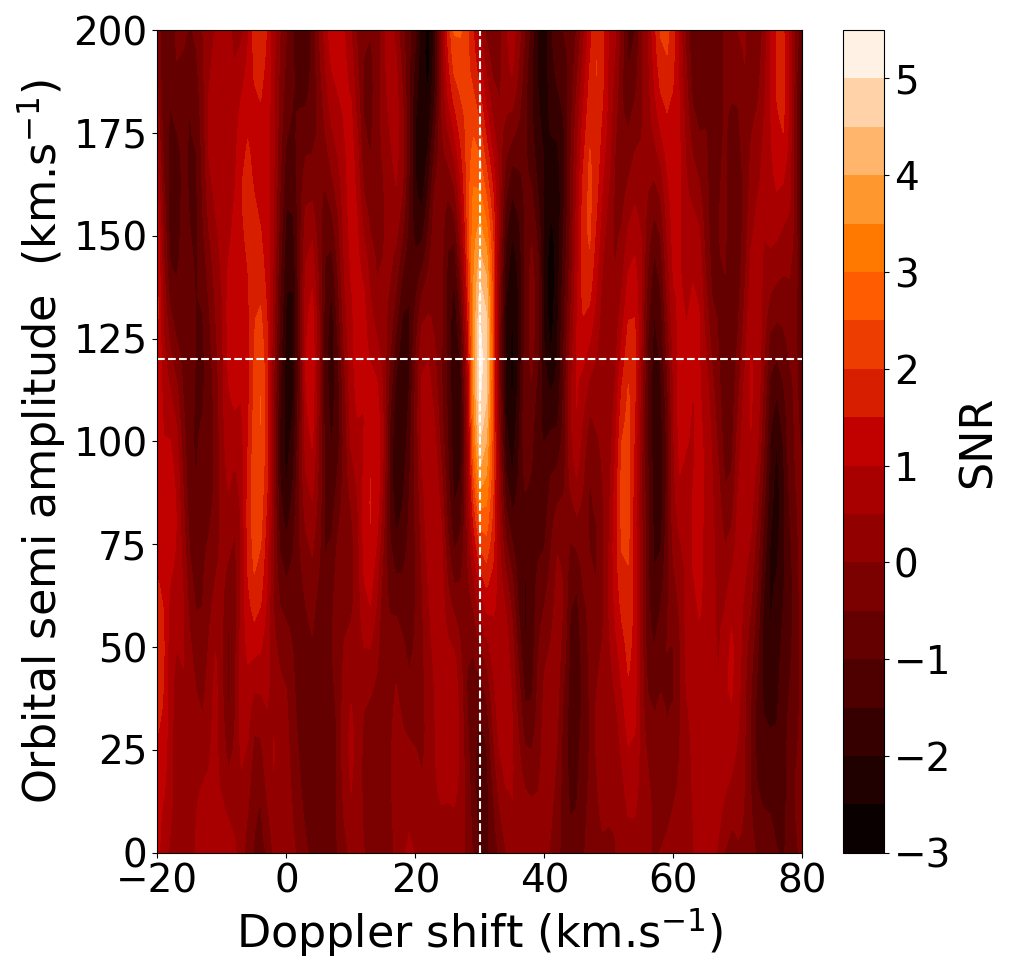}
    \includegraphics[width=0.46\linewidth]{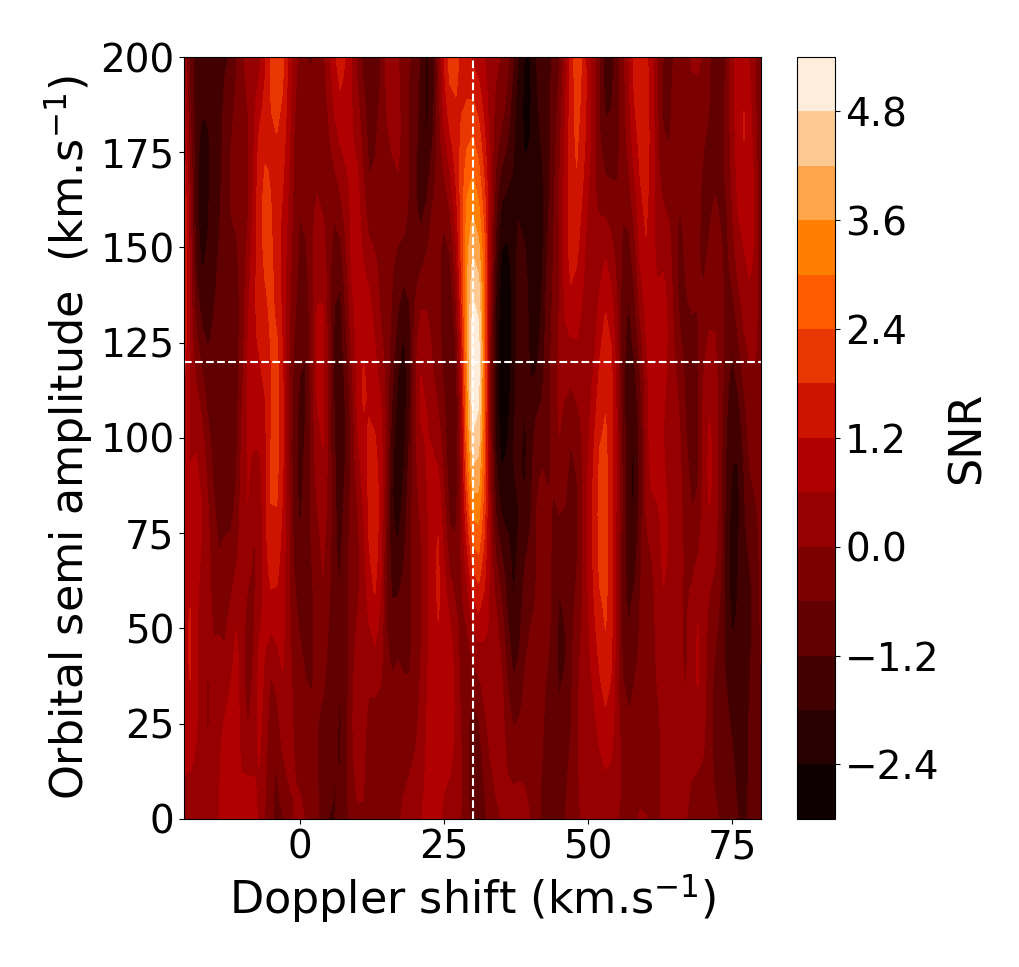}
    \caption{Cross correlation significance between synthetic data using Model 1 (top left), 2 (top right) and 3 (bottom) (see Section \ref{sec:Data})  and the same model with all molecules as a function of Doppler velocity and semi-amplitude}
    \label{fig:corr_tot}

\end{figure*}

\begin{figure}
    \includegraphics[width=\linewidth]{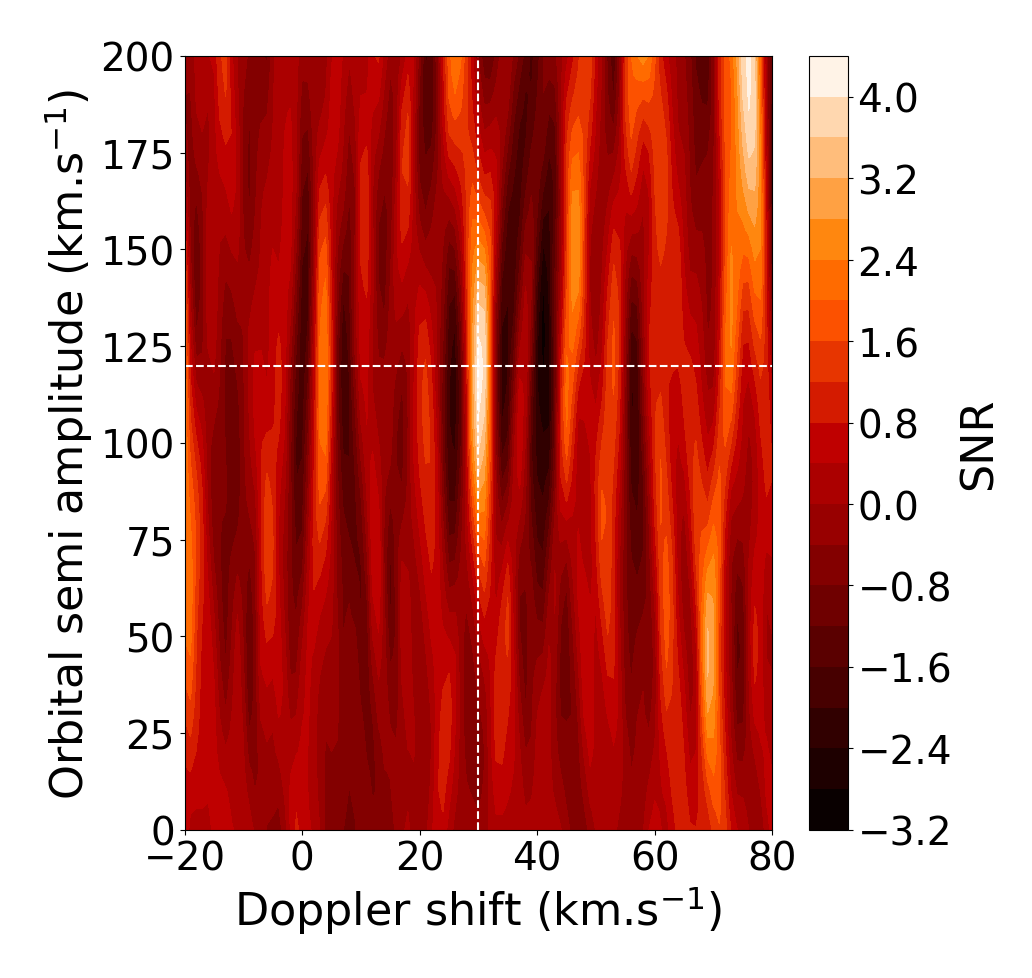}
    \caption{Cross correlation significance between synthetic data using Model 3 (see Section \ref{sec:Data}) and a model containing only CH$_4$ as a function of Doppler velocity and semi-amplitude}
    \label{fig:corr_CH4-3}
\end{figure}

\begin{figure}
    \includegraphics[width=\linewidth]{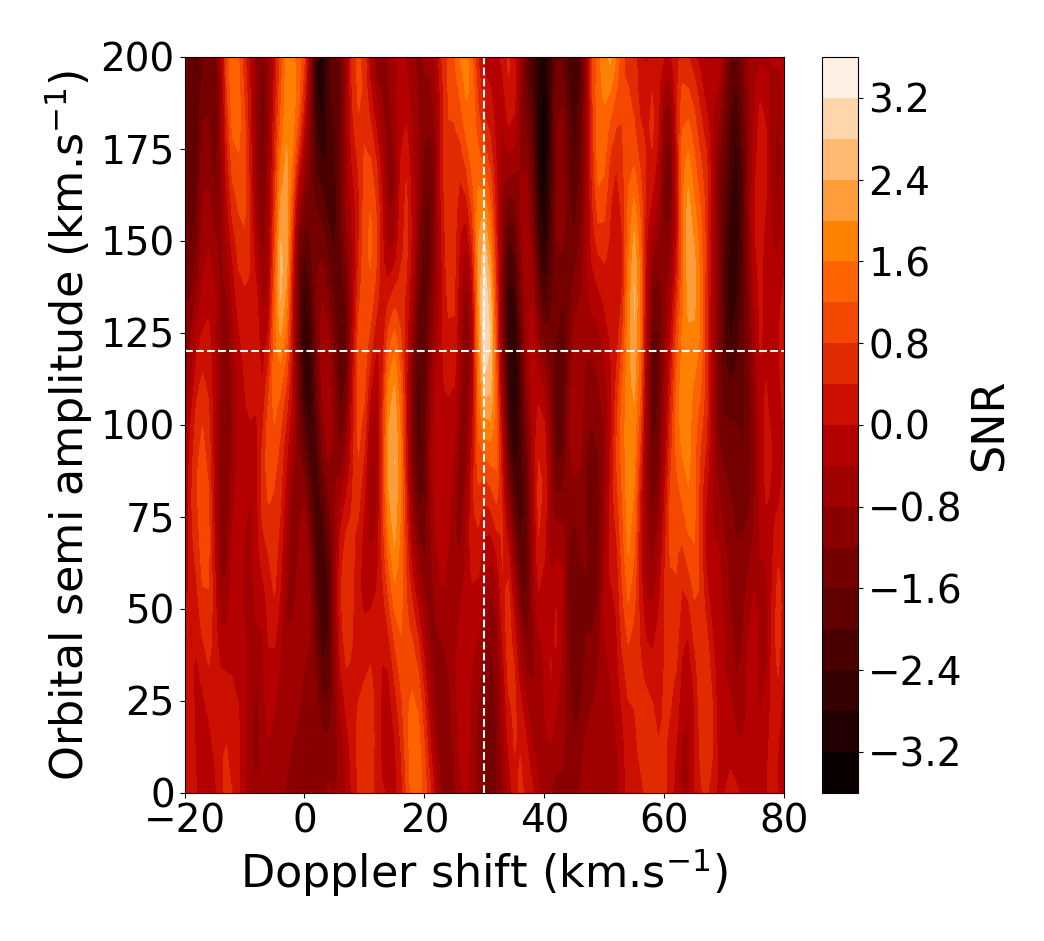}
    \includegraphics[width=\linewidth]{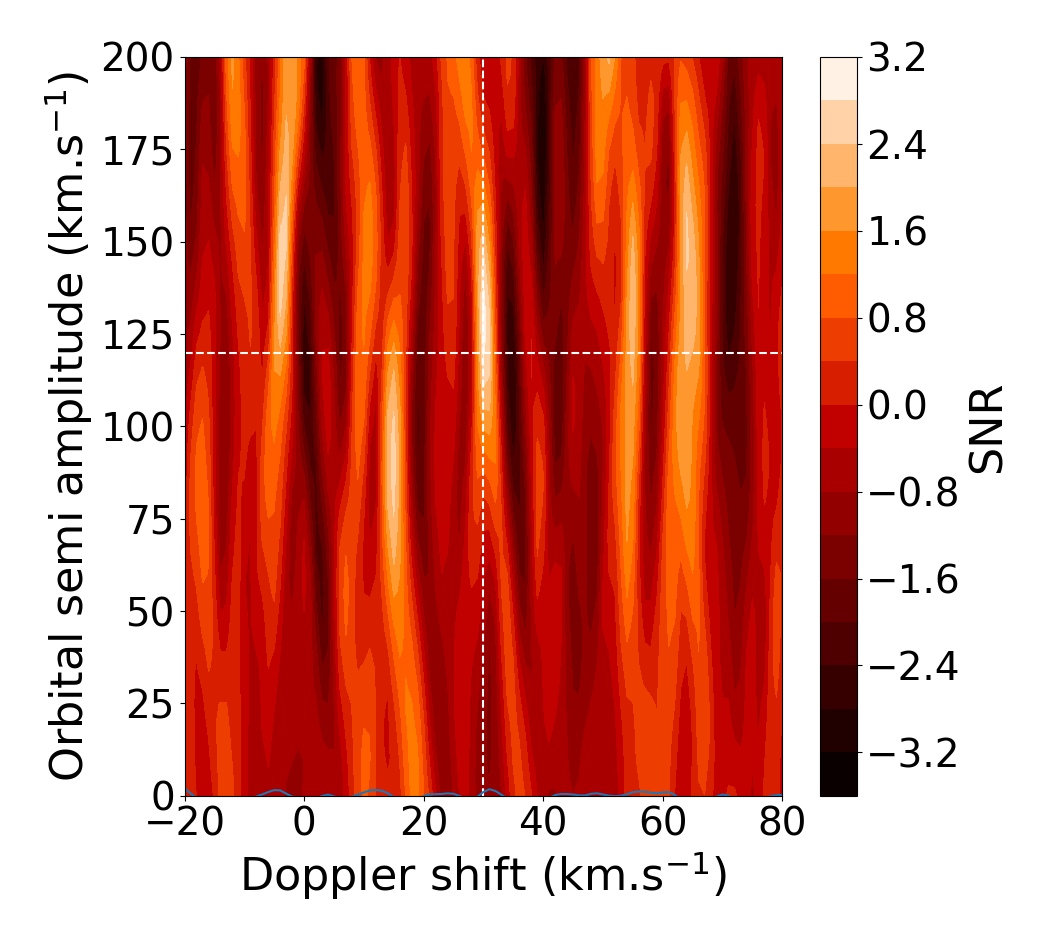}
    \caption{Cross correlation significance between synthetic data using Model 1(top, see Section \ref{sec:Data}) or Model 2 (bottom) and a model containing only NH$_3$ as a function of Doppler velocity and semi-amplitude}
    \label{fig:corr_NH3-12}
\end{figure}

\begin{figure}
    \includegraphics[width=\linewidth]{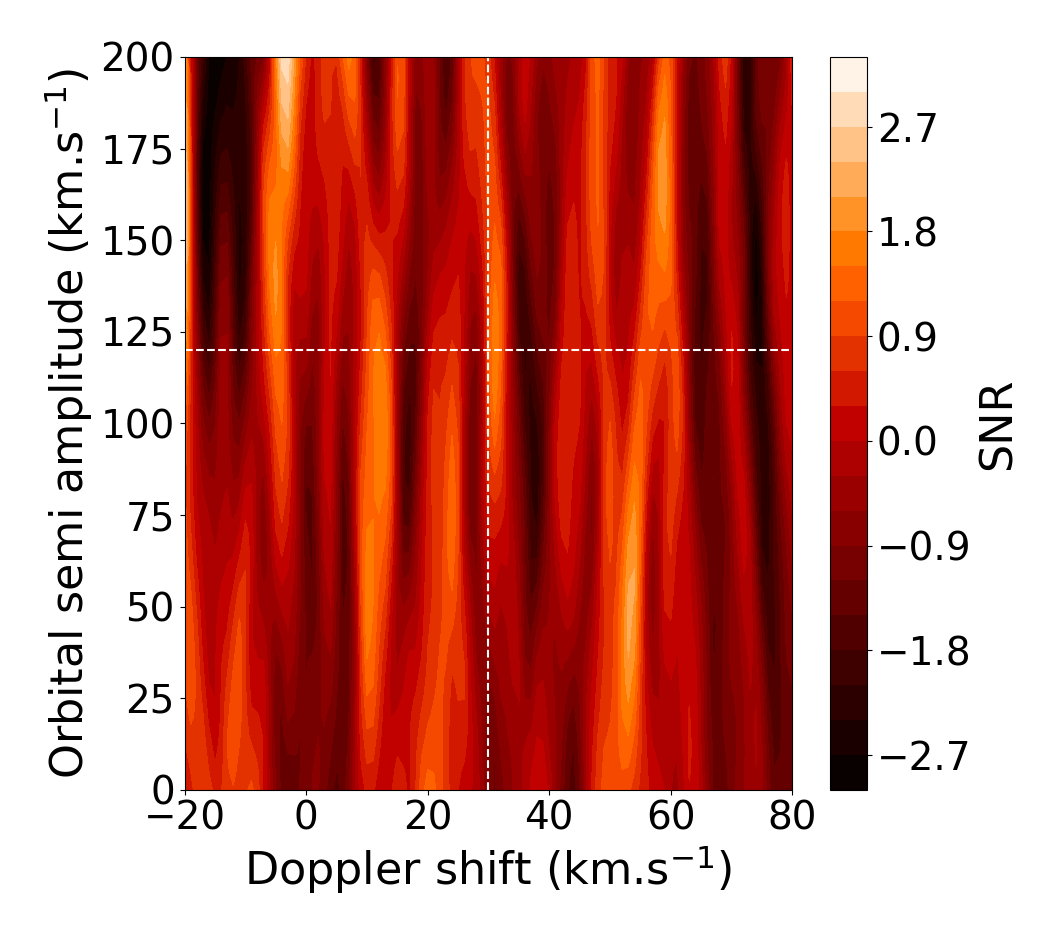}
    \includegraphics[width=\linewidth]{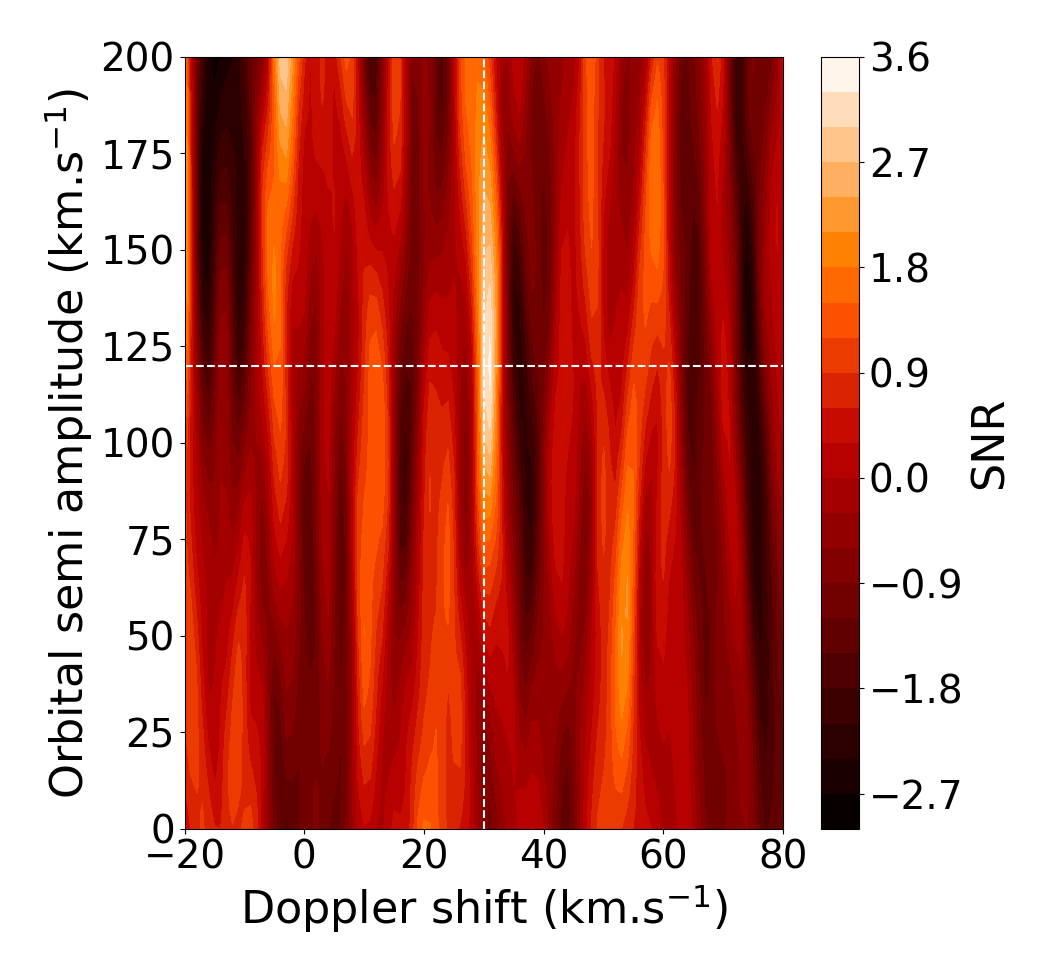}
    \caption{Cross correlation significance between synthetic data using Model 1(top, see Section \ref{sec:Data}) or Model 2 (bottom) and a model containing only H$_2$O as a function of Doppler velocity and semi-amplitude}
    \label{fig:corr_H2O-23}
\end{figure}

\newpage
\section{Multi species model - retrieval}
\label{app:marg}
\begin{figure*}
    \centering
    \includegraphics[width=\linewidth]{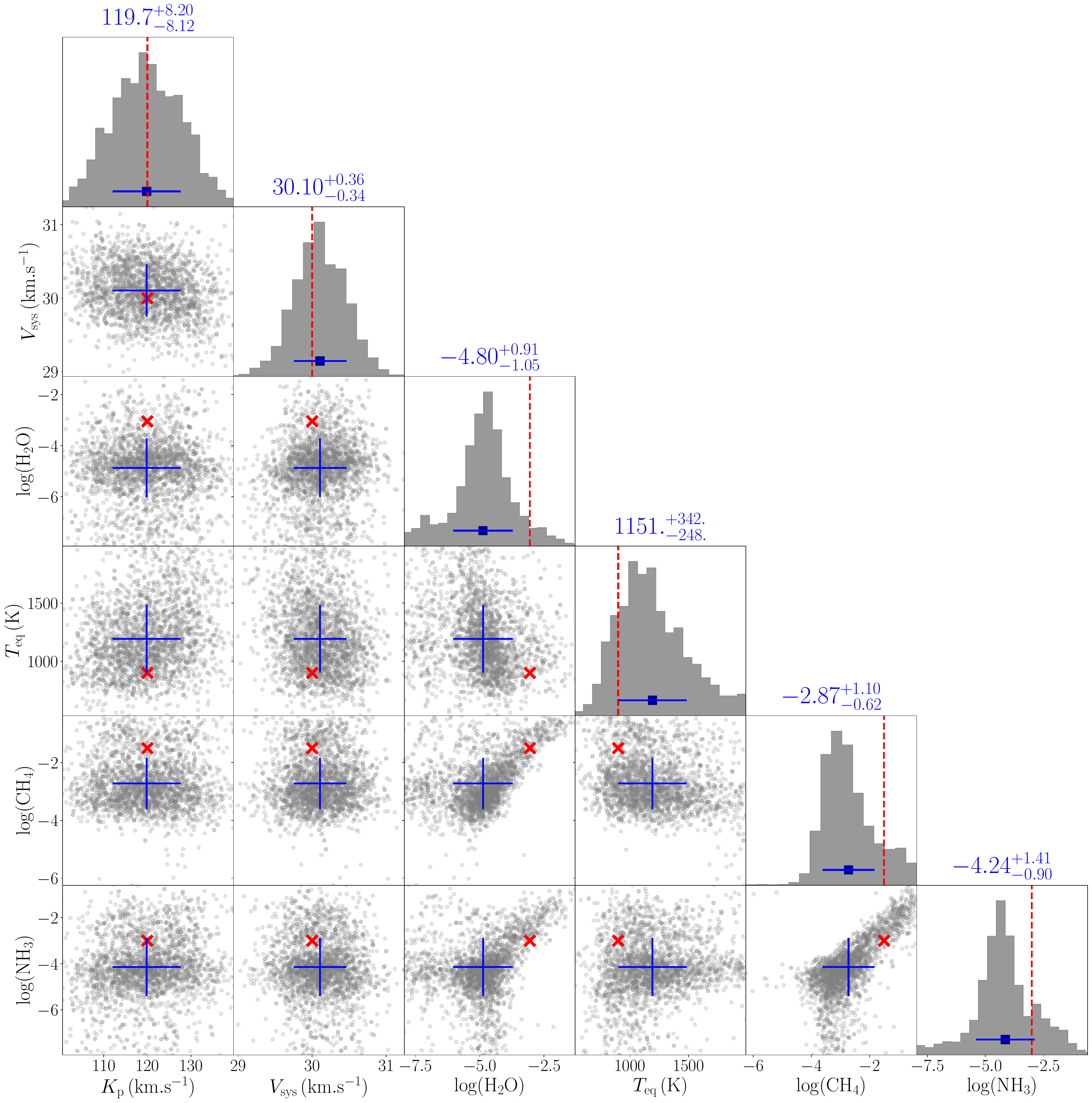}
    \caption{Corner plot of the result from a pymultinest retrieval with a model with multiple species, with VMR taken from the Model 1 of table \ref{tab:multi} (inspired by table 4 of the extended data of \citet{Giacobbe2021} as a fiducial example). The blue cross and bar shows the best fit value with the $1\sigma$ error bar, the red line and crosses show the injected values.}
    \label{fig:nested_multi}
\end{figure*}

\begin{figure*}
    \centering
    \includegraphics[width=\linewidth]{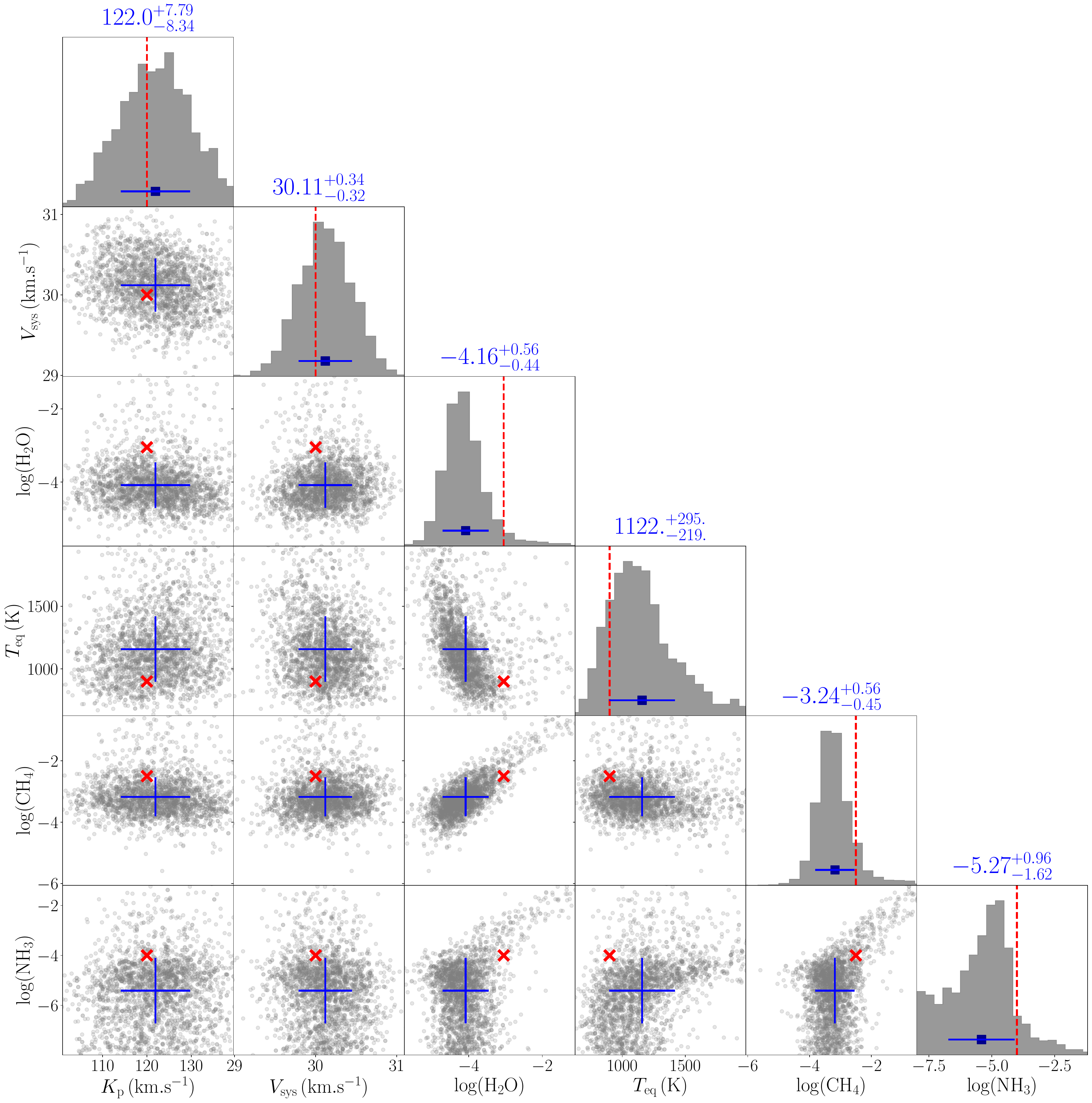}
    \caption{Same as Fig.\ref{fig:nested_multi} with Model 2 of Table \ref{tab:multi}}
    \label{fig:nested_multi_reduced}
\end{figure*}

\begin{figure*}
    \centering
    \includegraphics[width=\linewidth]{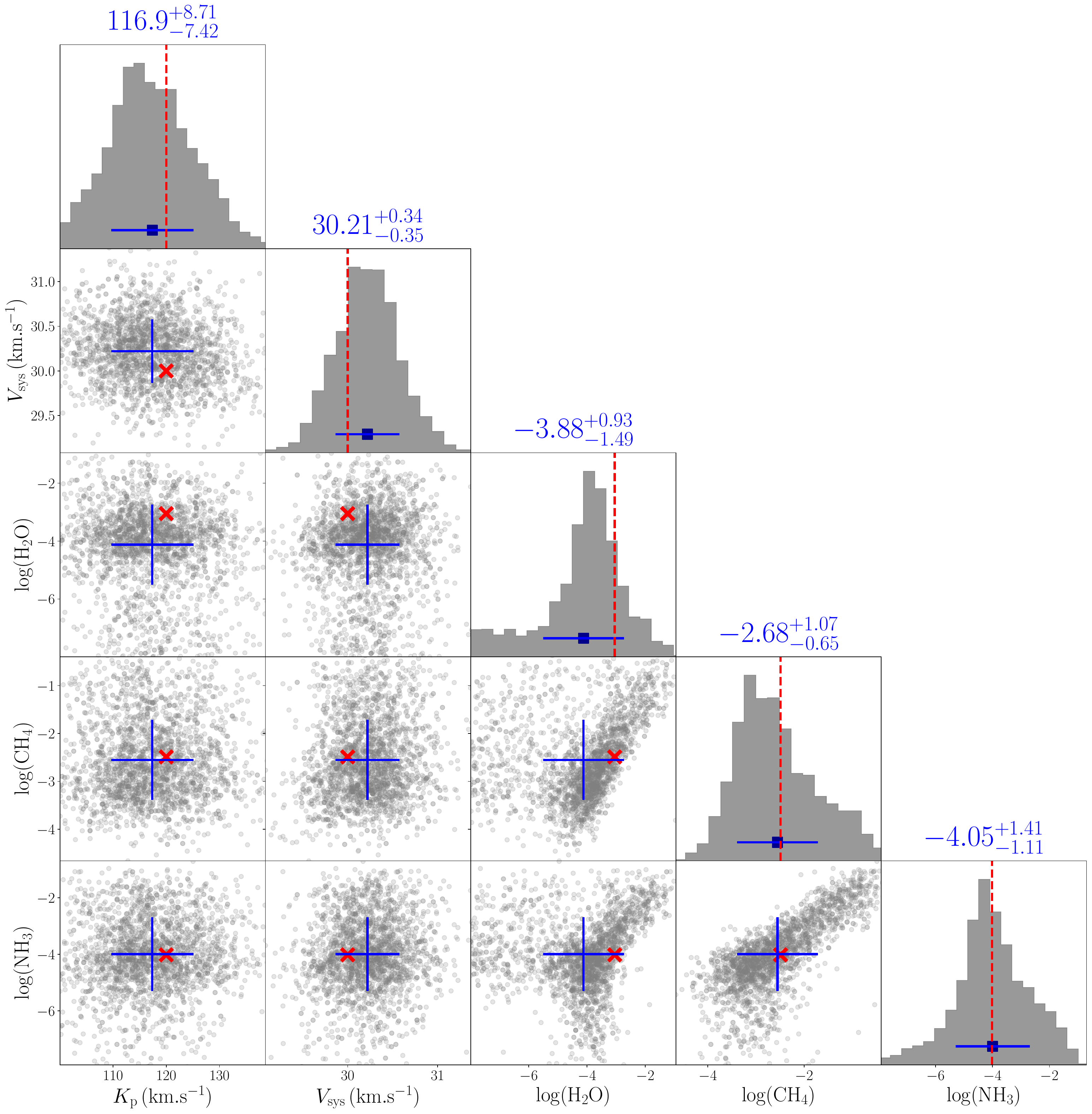}
    \caption{Same as Fig.\ref{fig:nested_multi} with Model 2 of Table \ref{tab:multi} and no temperature retrieval}
    \label{fig:nested_multi_reduced_fixT}
\end{figure*}

\begin{figure*}
    \centering
    \includegraphics[width=\linewidth]{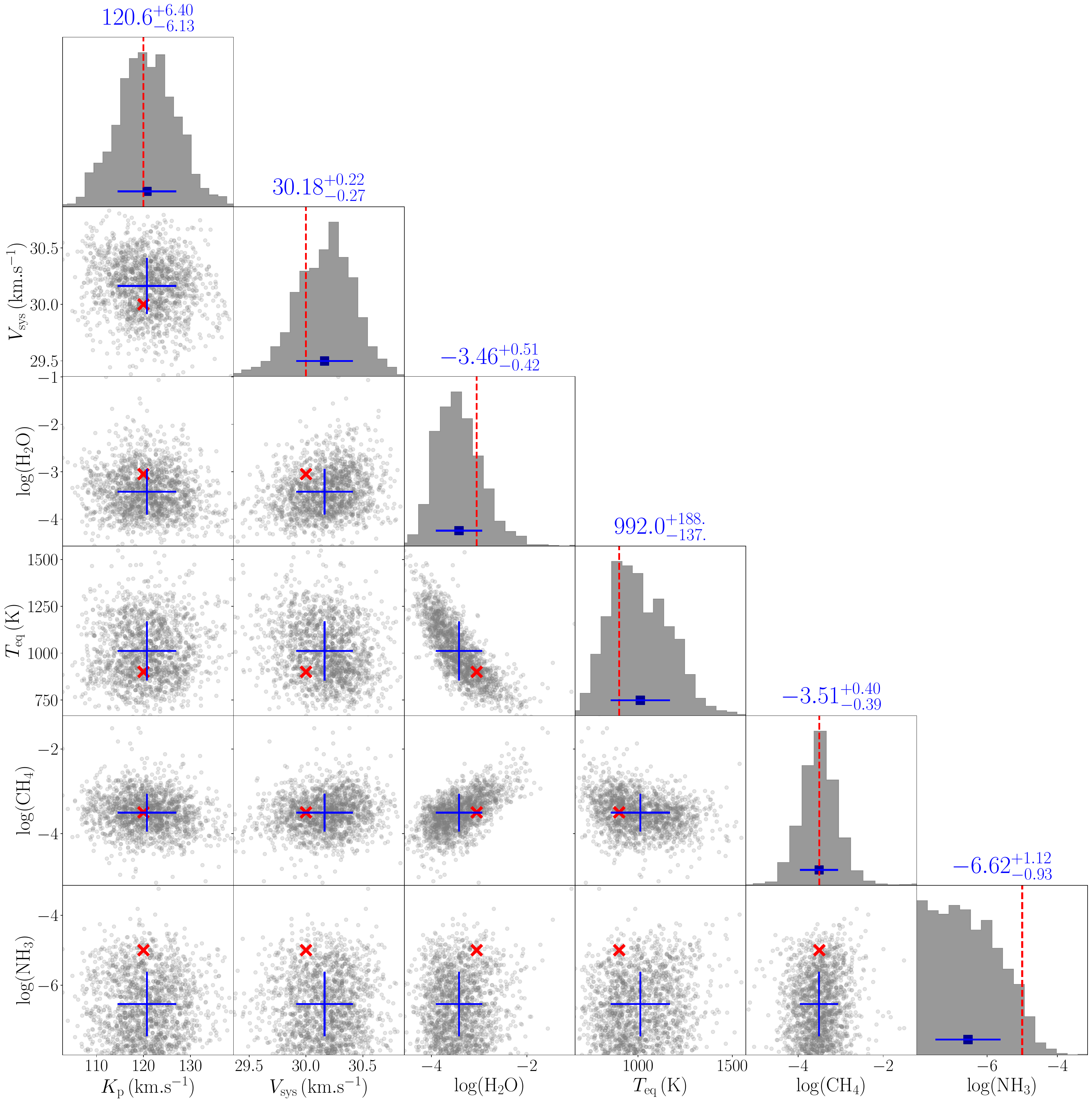}
    \caption{Same as Fig.\ref{fig:nested_multi} with Model 3 of Table \ref{tab:multi}. The shape of the NH$_3$ distribution confirms that this molecule is not detected in this model and we only obtain an upper limit for its content.}
    \label{fig:nested_multi_reduced100}
\end{figure*}

\begin{figure*}
    \centering
    \includegraphics[width=\linewidth]{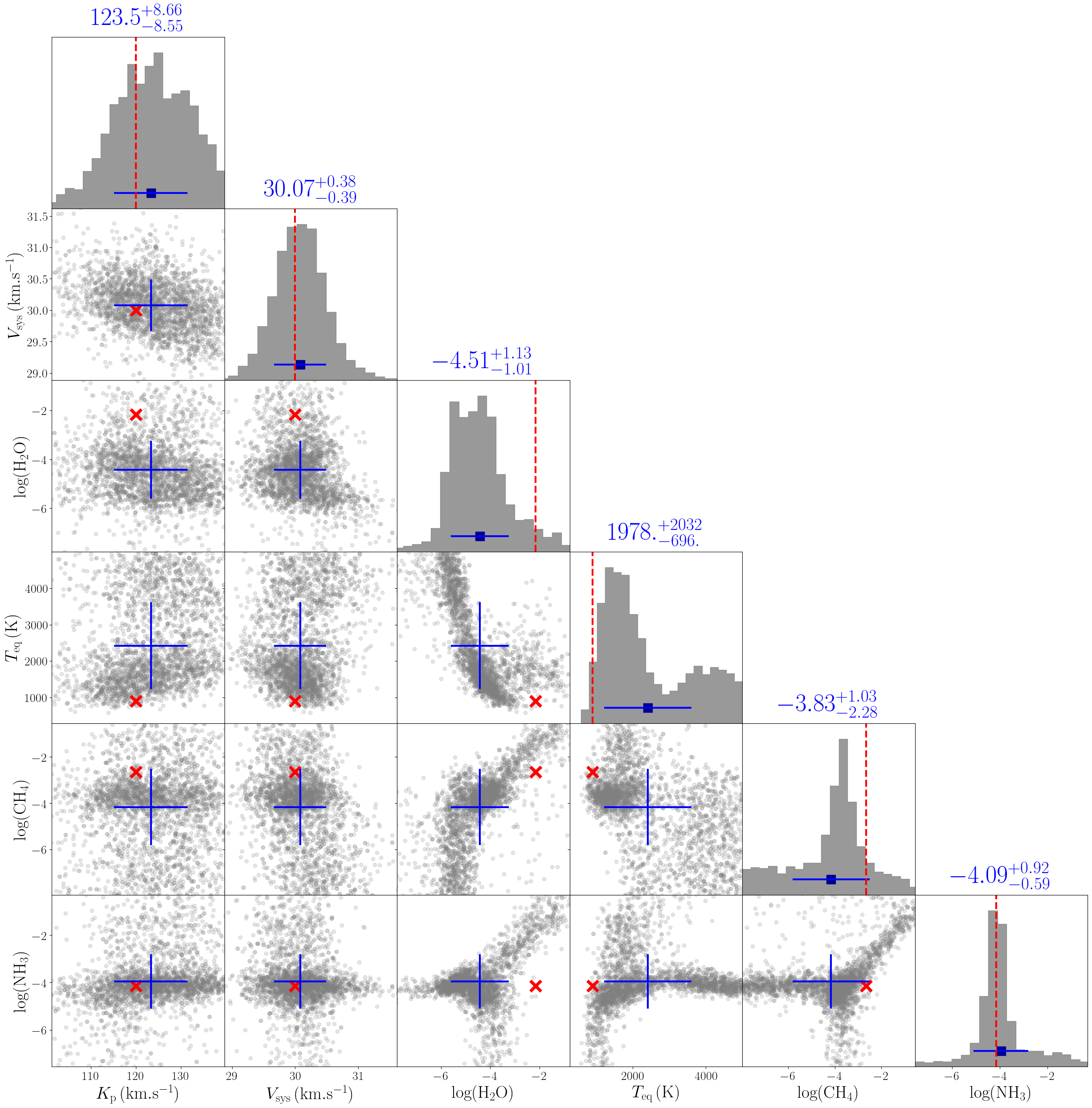}
    \caption{{Same as Fig.\ref{fig:nested_multi} with Model 3 of Table \ref{tab:multi} and removing orders 68, 52 and 39 (centered at 1.13, 1.47 and 1.96 microns). We recover a NH$_3$ detection, albeit with a very large temperature degeneracy as mentioned in the text}}
    \label{fig:nested_multi_reduced100_NH3}
\end{figure*}

\bsp	
\label{lastpage}
\end{document}